\documentclass[iop,onecolumn]{emulateapj}
\usepackage{graphicx,epsf}
\newcommand{\ltsim}{\protect\raisebox{-0.5ex}{$\:\stackrel{\textstyle <}{\sim}\:$}}
\newcommand{\gtsim}{\protect\raisebox{-0.5ex}{$\:\stackrel{\textstyle >}{\sim}\:$}}

\shorttitle{RMHD Simulations of Protostellar Core Formation}
\shortauthors{Tomida et al.}
\begin{document}
\title{RADIATION MAGNETOHYDRODYNAMIC SIMULATIONS OF PROTOSTELLAR COLLAPSE: PROTOSTELLAR CORE FORMATION}

\author{Kengo Tomida\altaffilmark{1,2,3}, Kohji Tomisaka\altaffilmark{2,3}, Tomoaki Matsumoto\altaffilmark{4}, Yasunori Hori\altaffilmark{3}, Satoshi Okuzumi\altaffilmark{5}, Masahiro N. Machida\altaffilmark{6}, and Kazuya Saigo\altaffilmark{3}}

\altaffiltext{1}{Department of Astrophysical Sciences, Princeton University, Princeton, NJ 08544, USA; \mbox{tomida@astro.princeton.edu}}
\altaffiltext{2}{Department of Astronomical Science, The Graduate University for Advanced Studies (SOKENDAI), Osawa, Mitaka, Tokyo 181-8588, Japan; \mbox{tomisaka@th.nao.ac.jp}}
\altaffiltext{3}{National Astronomical Observatory of Japan, Osawa, Mitaka, Tokyo 181-8588, Japan; \mbox{yasunori.hori@nao.ac.jp}, \mbox{saigo.kazuya@nao.ac.jp}}
\altaffiltext{4}{Faculty of Humanity and Environment, Hosei University, Fujimi, Chiyoda-ku, Tokyo 102-8160, Japan; \mbox{matsu@hosei.ac.jp}}
\altaffiltext{5}{Department of Physics, Nagoya University, Furo-cho, Chikusa-ku, Nagoya, Aichi 464-8602, Japan; \mbox{okuzumi@nagoya-u.jp}}
\altaffiltext{6}{Department of Earth and Planetary Sciences, Faculty of Sciences, Kyushu University, Hakozaki, Higashi-ku, Fukuoka 812-8581, Japan; \mbox{machida.masahiro.018@m.kyushu-u.ac.jp}}

\begin{abstract}
We report the first three-dimensional radiation magnetohydrodynamic (RMHD) simulations of protostellar collapse with and without Ohmic dissipation.We take into account many physical processes required to study star formation processes, including a realistic equation of state. We follow the evolution from molecular cloud cores until protostellar cores are formed with sufficiently high resolutions without introducing a sink particle. The physical processes involved in the simulations and adopted numerical methods are described in detail.We can calculate only about one year after the formation of the protostellar cores with our direct three-dimensional RMHD simulations because of the extremely short timescale in the deep interior of the formed protostellar cores, but successfully describe the early phase of star formation processes. The thermal evolution and the structure of the first and second (protostellar) cores are consistent with previous one-dimensional simulations using full radiation transfer, but differ considerably from preceding multi-dimensional studies with the barotropic approximation. The protostellar cores evolve virtually spherically symmetric in the ideal MHD models because of efficient angular momentum transport by magnetic fields, but Ohmic dissipation enables the formation of the circumstellar disks in the vicinity of the protostellar cores as in previous MHD studies with the barotropic approximation. The formed disks are still small (less than 0.35 AU) because we simulate only the earliest evolution. We also confirm that two different types of outflows are naturally launched by magnetic fields from the first cores and protostellar cores in the resistive MHD models.
\end{abstract}


\section{Introduction}
As stars are the most fundamental elements in the universe, their formation is one of the longstanding key topics in astrophysics. Because of its complicated nature and observational difficulties, computational simulations have played crucial roles in this field. \citet{lrs69} first showed the scenario from molecular cloud cores to protostellar cores with one dimensional hydrodynamic simulations using the diffusion approximation for radiation transfer, and found that two remarkable quasi-hydrostatic objects are formed in the course of protostellar collapse. A molecular cloud core initially collapses almost isothermally because thermal emission from dust grains is efficient in this phase. The gas temperature starts to rise when the central region gets dense and radiation cooling becomes inefficient, then the collapse almost stops and a quasi-hydrostatic object forms. This object, a so-called first (hydrostatic) core, evolves through accretion from the envelope. When the central temperature reaches about 2,000 K where hydrogen molecules start to dissociate, the core becomes unstable and collapses again because ${\rm H_2}$ dissociation is strongly endothermic and its gas pressure fails to balance gravity. This second collapse ends when molecular hydrogen dissociates completely and another quasi-static object is formed. This is the second core or the protostellar core. It acquires its mass via accretion on the dynamical timescale of the natal cloud core and evolves into a protostar and eventually a main-sequence star. More sophisticated simulations have been performed \citep{wn80,wn80b,sst1,sst2,sst3,mi00} and the scenario of protostellar collapse under spherical symmetry is by now well established.

However, there is non-negligible internal motion (i.e., rotation and turbulence) in molecular cloud cores which makes the collapse completely different from the spherically-symmetric models. Because centrifugal force is typically strong enough to prevent formation of stars (``the angular momentum problem"), there must be sufficiently effective mechanisms of angular momentum transport. Gravitational torque via non-axisymmetric structure like spiral arms takes this role if magnetic fields are very weak. \citet{bate98} first performed 3D Smoothed Particle Hydrodynamics (SPH) simulations and showed that gravitational torque redistributes angular momentum sufficiently and enables formation of protostellar cores. On the other hand, \citet{tmsk98,tmsk00,tmsk02} showed that magnetic fields transport angular momentum far more efficiently via torsional Alfv\'en waves (so-called magnetic braking, \citet{mp79,mp80}) and bipolar outflows using two-dimensional MHD simulations. \citet{mcd05a,mcd05b} performed three-dimensional nested-grid MHD simulations of similar situations and investigated the condition for fragmentation in the early phase before the second collapse begins. \citet{mim06} investigated the evolution after the second collapse and showed that two different outflows are launched from the first core and the protostellar core. More detailed 3D simulations on this problem have been performed \citep{bnj06,hf08,ht08,mtmi08} and effects of many physical ingredients such as misaligned magnetic fields \citep{mt04,mmht06,ch10,joos12}, metallicities \citep{machida08} and turbulence \citep{sei12,sl12}, have been investigated. Even weak magnetic fields can transport angular momentum very efficiently and modify the structure of the accretion flow significantly, and observations suggest that molecular clouds are considerably magnetized \citep{hc05,gir06,fal08,tc08,cht09}. Therefore, magnetic fields are likely to play dominant roles in typical present-day environments.

Non-ideal MHD effects are also important in protostellar collapse. Typical magnetic flux in a molecular cloud core is far larger compared to that in a formed star, which means that there must be significant redistribution of magnetic flux during collapse. Moreover, \citet{ml08,ml09} claimed that rotationally-supported circumstellar disks are difficult to form in magnetized molecular cloud cores because the angular momentum transport due to magnetic fields is too efficient (``the magnetic braking catastrophe", see also \citet{li11}). Non-ideal MHD effects such as Ohmic dissipation and ambipolar diffusion are supposed to be responsible in this process, because the ionization degree in star forming clouds is very low. Ohmic dissipation redistributes sufficient magnetic flux outward from the centrally-condensed high density region and enables formation of circumstellar disks in the vicinity of protostars \citep{db10,inu10,mim11b,mm11}. Gravitational torque becomes important in transporting angular momentum where magnetic fields are significantly weakened \citep{mim10}.

In those previous multi-dimensional simulations, they adopted the barotropic approximation for gas thermodynamics in which the thermal evolution is approximated with simple polytropic relations mimicking the results of 1D RHD simulations \citep[e.g.,][]{mi00}, instead of solving radiation transfer. Recently, \citet{bate10,bate11} \citep[see also][]{wb06} reported 3D SPH RHD simulations of formation of protostellar cores. \cite{sch11} also performed similar calculations of protostellar collapse using 2D axisymmetric RHD simulations. They commonly found interesting phenomena happen when the protostellar cores are formed; bipolar outflows are launched from the disk-like first cores via radiation heating (not radiation force) from the protostellar cores. Considering the energy released in the second collapse and subsequent accretion onto the protostellar core, the irradiation is sufficient to heat up the gas in the first core and launch the outflows \citep{bate11}. The barotropic approximation cannot reproduce such outflows driven by radiation heating. The structure of the first core which depends on the angular momentum distribution seems to be important in this phenomenon, and in the extreme case, the first core is almost blown up \citep{sch11}. Such violent phenomena may affect the story of protostellar collapse, circumstellar disk formation, and evolution of protostars.

Thus both magnetic fields and radiation transfer are critically important in star formation processes. \citet{com10} and \citet{tomida10a} independently performed 3D RMHD simulations of protostellar collapse, but both stopped their calculations before the onset of the second collapse and they adopted the ideal MHD approximation. In this work, we report the first 3D RMHD simulations of protostellar collapse from molecular cloud cores to protostellar cores with and without Ohmic dissipation. To follow the evolution after the second collapse, we adopt a realistic equation of state (EOS) which takes chemical reactions into account. The goal of this study is to describe the realistic evolution in the early phase of protostellar collapse (i.e., until protostellar cores are formed) involving both radiation transfer and magnetic fields.

The plan of this paper is as follows: We describe the basic equations in Section 2. The adopted numerical methods are given in Section 3 and the simulation setups in Section 4. We show the results of our 3D RMHD simulations in Section 5. Section 6 is devoted to conclusions and discussions. In the Appendices, we explain the microphysical processes considered in our simulations.

\section{Basic Equations}
In this work, we solve system equations including MHD, self-gravity, radiation transfer, a realistic EOS, and Ohmic dissipation. We use the comoving-frame RHD equations in which all the radiation related variables are defined in the comoving frame of the local fluid element \citep{castor}. For simplicity and to reduce computational load, we adopt the gray approximation (using frequency-averaged equations assuming the local thermodynamic equilibrium) and the flux limited diffusion approximation \citep[FLD,][]{lp81,lev84} for radiation transfer. We take Ohmic dissipation into account as the first step because it is supposed to be the most significant in the context of magnetic flux loss from high density gas \citep{mim07}, although other non-ideal MHD effects such as ambipolar diffusion and Hall effects are also important \citep{nkn02}. 

The governing equations are as follows:
\begin{eqnarray}
\frac{\partial\rho}{\partial t}+\nabla\cdot(\rho\mathbf{v})=0,\label{mc}\\
\frac{\partial\rho\mathbf{v}}{\partial t}+\nabla\cdot\left[\rho\mathbf{v}\otimes\mathbf{v}+\left(p+\frac{1}{2}|\mathbf{B}|^2\right)\mathbb{I}-\mathbf{B}\otimes\mathbf{B}\right]=-\rho\nabla\Phi+\frac{\sigma_R}{c}\mathbf{F}_r,\\
\frac{\partial\mathbf{B}}{\partial t}-\nabla\times(\mathbf{v}\times\mathbf{B}-\eta\nabla\times\mathbf{B})=0,\label{induction}\\
\nabla\cdot\mathbf{B}=0,\label{sole}\\
\frac{\partial e}{\partial t}+\nabla\cdot\left[\left(e+p+\frac{1}{2}|\mathbf{B}|^2\right)\mathbf{v}-\mathbf{B}(\mathbf{v}\cdot\mathbf{B})-\eta\mathbf{B}\times(\nabla\times\mathbf{B})\right]=\nonumber\\
-\rho\mathbf{v}\cdot\nabla\Phi-c\sigma_P(a_r T_g^4-E_r)+\frac{\sigma_R}{c}\mathbf{F}_r\cdot\mathbf{v},\label{energy}\\
\nabla^2\Phi=4\pi G\rho,\\
\frac{\partial E_r}{\partial t}+\nabla\cdot[\mathbf{v}E_r]+\nabla\cdot\mathbf{F}_r+\mathbb{P}_r:\nabla\mathbf{v}=c\sigma_P(a_r T_g^4 -E_r),\\
\mathbf{F}_r=\frac{c\lambda}{\sigma_R}\nabla E_r,\hspace{1em}
\lambda(R)=\frac{2+R}{6+2R+R^2},\hspace{1em}
R=\frac{|\nabla E_r|}{\sigma_R E_r},\nonumber\\
\mathbb{P}_r=\mathbb{D}E_r,\hspace{1em}
\mathbb{D}=\frac{1-\chi}{2}\mathbb{I}+\frac{3\chi-1}{2}\mathbf{n}\otimes \mathbf{n},\hspace{1em}
\chi=\lambda+\lambda^2R^2,\hspace{1em}
\mathbf{n}=\frac{\nabla E_r}{|\nabla E_r|}\nonumber,\label{end}
\end{eqnarray}
where $\rho$ denotes the gas density, $\mathbf{v}$ the fluid velocity, $\mathbf{B}$ the magnetic flux density, $p$ the gas pressure, $T_g$ the gas temperature, $e=e_g+\frac{1}{2}\rho v^2+\frac{1}{2}|\mathbf{B}|^2$ the total gas energy density ($e_g$ is the internal energy density of the gas), $E_r$ the radiation energy density, $\Phi$ the gravitational potential, $\mathbf{F}_r$ the radiation energy flux, and $\mathbb{P}_r$ the radiation pressure tensor, respectively. $c=2.99792 \times 10^{10} \, {\rm cm\,s^{-1}}$ is the speed of light and $G= 6.673 \times 10^{-8} \, {\rm cm^3\,g^{-1}\,s^{-2}}$ is the gravitational constant. $a_r=4\sigma/c=7.5657 \times 10^{-15}{\rm erg \, cm^{-3} \, K^{-4}}$ is the radiation (density) constant where $\sigma=5.6704 \times 10^{-5} {\rm erg \, cm^{-2} \, K^{-4}}$ is the Stefan-Boltzmann constant. $\eta$ is the resistivity and $\sigma_R$ ($\sigma_P$) is the Rosseland (Planck) mean opacity. $\mathbb{I}$ denotes the unit matrix. ``:" means the double dot product of two tensors, $\mathbb{A}:\mathbb{B}=\sum_i\sum_j A_{ij}B_{ji}$. These equations represent conservation of mass, the equation of motion, the induction equation including Ohmic dissipation, the solenoidal constraint, the Poisson's equation of gravity, and the gray FLD radiation transfer equations from top to bottom. Basically we use the Gaussian cgs units but we rescale the magnetic flux density to eliminate the constant coefficients, i.e., $\mathbf{B}=\mathbf{B}_0/\sqrt{4\pi}$ where $\mathbf{B}_0$ is given in Gauss. Additionally, we need an EOS which gives the relations between the thermodynamic variables $\rho, p, T$ and $e_g$ to close the system. We adopt the tabulated EOS in which we consider internal degrees of freedom and chemical reactions of seven species (${\rm H_2, H, H^+, He, He^+, He^{2+}}$ and ${\rm e^-}$). Here we assume the solar abundance, $X=0.7$ and $Y=0.28$. We also adopt the tabulated tables of the Rosseland and Planck mean opacities. In order to cover the huge dynamic range, we combine three opacity tables: \citet{semenov}, \citet{op94} and \citet{fer05}. The resistivity is adopted from \citet{un09}, while we additionally consider thermal ionization of potassium (K). The detailed descriptions about these microphysical processes are given in Appendices.

Note that \citet{mim06,mim07,mtmi08} simplified the resistive term in eq.~\ref{induction} to $\eta\nabla^2\mathbf{B}$, but it is inadequate because the resistivity $\eta$ has large spatial gradient according to the local gas density and temperature gradients, and what is worse, it violates the divergence free condition (eq.~\ref{sole}). In this work we correctly discretize the original equation, and also includes the effect of the resistivity in the energy equation (eq.~\ref{energy}).

\section{Method}
\subsection{Operator-Splitting\label{os}}
In order to solve the complex system described in the previous section, we divide the system into five parts and solve them separately on the nested-grid hierarchy. We update our system in the following strategy:

\noindent{\bf Step 1. Ideal MHD part:}
\begin{eqnarray}
\frac{\partial\rho}{\partial t}+\nabla\cdot(\rho\mathbf{v})&=&0,\label{mhds}\\
\frac{\partial\rho\mathbf{v}}{\partial t}+\nabla\cdot\left[\rho\mathbf{v}\otimes\mathbf{v}+\left(p+\frac{1}{2}|\mathbf{B}|^2\right)\mathbb{I}-\mathbf{B}\otimes\mathbf{B}\right]&=&0,\\
\frac{\partial\mathbf{B}}{\partial t}-\nabla\times(\mathbf{v}\times\mathbf{B})&=&0,\\
\nabla\cdot\mathbf{B}&=&0,\\
\frac{\partial e}{\partial t}+\nabla\cdot\left[\left(e+p+\frac{1}{2}|\mathbf{B}|^2\right)\mathbf{v}-\mathbf{B}(\mathbf{v}\cdot\mathbf{B})\right]&=&0,\label{mhde}\\
\frac{\partial E_r}{\partial t}+\nabla\cdot[\mathbf{v}E_r]&=&0. \label{advf}
\end{eqnarray}
{\bf Step 2. Self-Gravity part:}
\begin{eqnarray}
\nabla^2\Phi&=&4\pi G\rho,\label{poisson}\\
\frac{\partial\rho\mathbf{v}}{\partial t}&=&-\rho\nabla\Phi,\label{grv1}\\
\frac{\partial e}{\partial t}&=&-\rho\mathbf{v}\cdot\nabla\Phi,\label{grv2}
\end{eqnarray}
{\bf Step 3. Resistivity part:}
\begin{eqnarray}
\frac{\partial\mathbf{B}}{\partial t}+\nabla\times(\eta\nabla\times\mathbf{B})&=&0, \label{md1}\\
\frac{\partial e}{\partial t}-\nabla\cdot\left[\eta\mathbf{B}\times(\nabla\times\mathbf{B})\right]&=&0.\label{md2}
\end{eqnarray}
{\bf Step 4. Radiation part:}
\begin{eqnarray}
\frac{\partial e_g}{\partial t}&=&-c\sigma_P(a_rT_g^4-E_r),\label{flds}\\
\frac{\partial E_r}{\partial t}+\nabla\cdot\mathbf{F}_r+\mathbb{P}_r:\nabla\mathbf{v}&=&c\sigma_P (a_r T_g^4 -E_r),\label{fld2}
\end{eqnarray}
{\bf Step 5. Radiation Force part:}
\begin{eqnarray}
\frac{\partial\rho\mathbf{v}}{\partial t}&=&\frac{\sigma_R}{c}\mathbf{F}_r,\\
\frac{\partial e}{\partial t}&=&\frac{\sigma_R}{c}\mathbf{F}_r\cdot\mathbf{v}.
\end{eqnarray}

Introducing such an operator-splitting technique causes loss of time-accuracy, and the overall accuracy of our time-integration scheme is first order.

\subsection{Magnetohydrodynamics and Self-Gravity}
We solve the MHD system using the HLLD approximate Riemann solver \citep{miyoshi}. It is suitable for simulations involving the realistic EOS because this scheme does not require the detailed knowledge about the EOS, unlike Roe's approximate Riemann solver. It is simple, robust, highly efficient, and almost as accurate as Roe's solver. We define all the variables at the cell center. To numerically satisfy the solenoidal constraint (\ref{sole}), we adopt the mixed correction based on the generalized Lagrangian multiplier approach proposed by \citet{dedner}. To achieve second-order accuracy in space and time, we adopt standard MUSCL (Monotone Upstream-centered Scheme for Conservation Laws) approach and the directionally-unsplit two-step predictor-corrector scheme \citep[e.g.,][]{hirsch}. We store ($\rho$, $\mathbf{v}$, $\mathbf{B}$, $e_g$, $\psi$, $E_r$) as primitive variables and interpolate them for spatial reconstruction with the minmod slope limiter for robustness ($\psi$ is an additional variable related to the solenoidal constraint). We use the gas internal energy $e_g$ as the primitive variable instead of the gas pressure $p$ which is the textbook notation in order to minimize the numerical error when we use the tabulated EOS. That is, the discretization error of the EOS table will be directly reflected in the results if we perform mutual conversions between $p$ and $e_g$ (i.e., $p\rightarrow e_g\rightarrow p$). We can avoid this error if we use one-way conversion ($e_g \rightarrow p$) only; the discretization error only affects the flux and its effect is small.

Some high resolution (magneto-) hydrodynamic solvers show strange oscillations at strong shocks aligned to grid structure. This problem called the Carbuncle phenomenon is obviously unphysical. To suppress this unphysical oscillation, we use the HLLD$-$ flux developed by \citet{hlldm} in the cells which potentially contain shocks. We adopt the method proposed in \citet{hmm} to identify such cells. HLLD$-$ is a modified version of HLLD, in which the tangential velocities ($v_y, v_z$) in the Riemann fan are replaced by the HLL averages while other variables are kept the same as HLLD. Sufficient but not too large additional shear viscosity is introduced by this procedure and it stabilizes the Carbuncle phenomenon.

We update the advective flux of radiation energy (\ref{advf}) separately using Roe's upwind method. When the radiation energy dominates the gas energy, the radiation pressure affects the sonic speed of the gas and the characteristics are modified, but we neglect this effect. In our star formation simulations (at least in low mass cases), we expect that we encounter such a radiation dominant region only in the deep interior of the protostellar core in the late phase of protostellar collapse, and we do not expect this effect will be significant.

To solve the Poisson's equation (\ref{poisson}) on the nested-grid hierarchy, we adopt the multigrid solver developed by \citet{mh03}. The solver gives second-order accurate solution in space. We integrate (\ref{grv1}) and (\ref{grv2}) using obtained gravitational potential.

\subsection{Ohmic Dissipation}
Because of the low ionization degree, non-ideal MHD effects such as Ohmic dissipation, ambipolar diffusion and the Hall effect work during protostellar collapse. Ohmic dissipation becomes dominant in the high density region (typically $\rho \gtsim 10^{-11}\, {\rm g\, cm^{-3}}$) and ambipolar diffusion dominates in the low density region \citep{nkn02,kunz09,kunz10}, but in this work we focus on Ohmic dissipation as the first step because it is supposed to be most significant in the context of the magnetic flux problem and the angular momentum problem \citep{mim11b,dapp12}. Taking account of ambipolar diffusion will be a future work. The timescale of Ohmic dissipation becomes shorter than the dynamical timescale in the high density region ($\rho \gtsim 10^{-10} \,  \rm{g \, cm^{-3}}$, Figure~\ref{resist}) and significant diffusion of magnetic flux occurs \citep{nkn02}. Note that these non-ideal MHD processes do not reduce total magnetic flux but redistribute centrally-concentrated magnetic flux outward from the high density region (see below, Figure~\ref{fmr}), although this process is often mentioned as "magnetic flux loss".

We discretize (\ref{md1}) and (\ref{md2}) in the same way as \citet{mat11}. Additionally, we introduce the correction terms proportional to $\nabla\cdot\mathbf{B}$ to improve the nature of the system as described in \citet{grv08}. Since the timescale of the diffusion of magnetic fields can be far shorter than the dynamical timescale (Figure~\ref{resist}), here we adopt the Super-Time-Stepping (STS) method proposed by \citet{sts} (for astrophysical applications, see \citet{osd06,choi09,com11}) when the timestep for the resistivity part is shorter than the hydrodynamic timestep. STS is constructed based on the simple explicit scheme, but we can relax the Courant-Friedrich-Levy (CFL) stability condition by using the following sequence of sub-timesteps:
\begin{eqnarray}
\tau_j=\Delta t_{exp}\left[(-1+\nu)\cos\left(\frac{2j-1}{N}\frac{\pi}{2}\right)+1+\nu\right]^{-1}.\label{sts1}
\end{eqnarray}
The longer time integration of $\Delta T_{STS}$ is achieved after $N$ sub-timesteps,
\begin{eqnarray}
\Delta T_{STS}=\sum_j^{N}\tau_j=\Delta t_{exp}\frac{N}{2\sqrt{\nu}}\left[\frac{(1+\sqrt{\nu})^{2N}-(1-\sqrt{\nu})^{2N}}{(1+\sqrt{\nu})^{2N}+(1-\sqrt{\nu})^{2N}}\right],\label{sts2}
\end{eqnarray}
where $\Delta t_{exp}$ is the explicit timestep which satisfies the CFL condition and $\nu$ is a small positive parameter which controls the stability and efficiency of the scheme. Smaller $\nu$ gives better acceleration but STS may give unphysical results when it is too small. The optimal choice of $\nu$ depends on the problem, but typically $\nu \sim 0.01$ seems to be good for parabolic problems. From (\ref{sts2}), the maximum acceleration for given $\nu$ compared to the explicit scheme ($\Delta T_{exp}=N\Delta t_{exp}$) is $0.5/\sqrt{\nu}$ and the most efficient calculation is realized around $N\sim 0.5/\sqrt{\nu}$. In this work, we adopt $\nu = 0.01$ and $N=6$. We apply this scheme repeatedly until required time update is achieved. Even in the most time-consuming situation (i.e., the magnetic Reynolds number is very low, $\rho \sim 10^{-9} \, {\rm g\, cm^{-3}}$) in our production runs, STS keeps the computational cost lower than or similar to the total of all the other parts (including MHD, self-gravity and radiation transfer) at most.

\subsection{Radiation Transfer}
The timescale related to radiation can be far shorter than that related to hydrodynamics in star formation simulations. The lightspeed $c=2.99792 \times 10^{10} \, {\rm cm\,s^{-1}}$ is far larger than typical fluid velocity in star formation, which is on the order of ($1$ -- $100) \times 10^5 \, {\rm cm \, s^{-1}}$. Moreover, the radiation source terms are strongly non-linear. Therefore the differential equations of the radiation subsystem can be very stiff. It is unreasonable to calculate such a system using explicit schemes or even STS. Therefore we adopt an implicit time-integration scheme which is stable regardless of the timescale of involved physical processes.

\subsubsection{Discretization}
We adopt the first-order backward Euler method which is simple and stable. Here we discretize (\ref{flds}) and (\ref{fld2}) in one-dimension (extension to multi-dimension is straightforward) as follows:
\begin{eqnarray}
\frac{e_{g,i}^{n+1}-e_{g,i}^n}{\Delta t}&=&-c\sigma_P^n \left[a_r \left(T_{g,i}^{n+1}\right)^4-E_{r,i}^{n+1}\right],\label{rhdd1}\\
\frac{E_{r,i}^{n+1}-E_{r,i}^n}{\Delta t}&-&\frac{1}{\Delta x}\left[\left(\frac{c\lambda}{\sigma_R}\right)_{i+\frac{1}{2}}^n \frac{E_{r,i+1}^{n+1}-E_{r,i}^{n+1}}{\Delta x}-\left(\frac{c\lambda}{\sigma_R}\right)_{i-\frac{1}{2}}^n \frac{E_{r,i}^{n+1}-E_{r,i-1}^{n+1}}{\Delta x}\right]\nonumber \\
&=&c\sigma_P^n \left[a_r \left(T_{g,i}^{n+1}\right)^4-E_{r,i}^{n+1}\right]-\mathbb{P}_{r,i}^{n+1}:(\nabla\mathbf{v})_i.\label{rhdd2}
\end{eqnarray}
Superscripts and subscripts denote the indices of the discretized time and space, respectively. To avoid numerical difficulties due to strong non-linearity in the flux limiter and the opacities and to achieve stable integration, we adopt the time-lagged opacities and flux limiter \citep{castor}. This may cause loss of the time accuracy of the scheme, but we confirmed that it does not matter in our production runs because the timescale of the evolution of radiation fields is similar to the hydrodynamic timescale and therefore well resolved. We also use the time-lagged Eddington Tensor $\mathbb{D}^n$, which yields the radiation energy tensor $\mathbb{P}_r^{n+1}=\mathbb{D}^n E_r^{n+1}$.

There are some ways to evaluate the opacities and the flux limiter at the cell interfaces, $i+\frac{1}{2}$. Here we follow \citet{hg03} and adopt the surface formula which gives good flux even at a sharp surface of optically thick material like the surface of a first core.
\begin{eqnarray}
\sigma_{R,i+\frac{1}{2}}=\min\left[\frac{\sigma_{R,i}+\sigma_{R,i+1}}{2},\max{\left(\frac{2\sigma_{R,i}\sigma_{R,i+1}}{\sigma_{R,i}+\sigma_{R,i+1}},\frac{4}{3\Delta x}\right)}\right].
\end{eqnarray}
We evaluate the flux limiter at the cell interface $\lambda(R)_{i+\frac{1}{2}}$ using the following equations:
\begin{eqnarray}
R_{i+\frac{1}{2}}=\frac{|(\nabla E_r)_{i+\frac{1}{2}}|}{\sigma_{R,i+\frac{1}{2}}{E_{r,i+\frac{1}{2}}}},\\
E_{r,i+\frac{1}{2}}=\frac{E_{r,i}+E_{r,i+1}}{2},\\
(\nabla E_r)_{i+\frac{1}{2}}=\frac{E_{r,i+1}-E_{r,i}}{\Delta x}.
\end{eqnarray}
In three dimensions, the radiation energy gradient should be replaced with
\begin{eqnarray}
(\nabla E_r)_{i+\frac{1}{2},j,k}&=&\left(\begin{array}{c}
\displaystyle \frac{E_{r,i+1}-E_{r,i}}{\Delta x}\\
\displaystyle \frac{E_{r,i+1,j+1,k}+E_{r,i,j+1,k}-E_{r,i+1,j-1,k}-E_{r,i,j-1,k}}{4\Delta y}\\
\displaystyle \frac{E_{r,i+1,j,k+1}+E_{r,i,j,k+1}-E_{r,i+1,j,k-1}-E_{r,i,j,k-1}}{4\Delta z}
\end{array}
\right).
\end{eqnarray}

\subsubsection{Newton-Raphson Iterations}
To solve the non-linear system (\ref{rhdd1}) and (\ref{rhdd2}), we perform the Newton-Raphson iterations \citep{nr}. In this method, we search for the zero-points of the residual functions $f_i(\mathbf{X})$. We can find the root iteratively using the following matrix equation based on the Taylor expansion:
\begin{eqnarray}
\sum_{j=1}^N\frac{\partial f_i}{\partial x_j}\delta X_j=-f_i(\mathbf{X}) \label{nrm}
\end{eqnarray}
In our system, $\mathbf{X}$ is the vector of the gas and radiation energy densities in all the cells: 
\begin{eqnarray}
\mathbf{X}=(e_{g,1},...,e_{g,i},e_{g,i+1},...,E_{r,1},...,E_{r,i},E_{r,i+1},...)^T.
\end{eqnarray}
The residual functions are given from (\ref{rhdd1}) and (\ref{rhdd2}) as follows:
\begin{eqnarray}
f^g_i&=&e_{g,i}^{n+1}-e_{g,i}^n+\Delta t c\sigma_P^n \left[a_r \left\{T_g(\rho_i,e_{g,i}^{n+1})\right\}^4-E_{r,i}^{n+1}\right],\\
f^r_i&=&E_{r,i}^{n+1}-E_{r,i}^n+\Delta t\Bigl[\frac{1}{\Delta x}\left\{\left(\frac{c\lambda}{\sigma_R}\right)_{i+\frac{1}{2}}^n\frac{E_{r,i+1}^{n+1}-E_{r,i}^{n+1}}{\Delta x}-\left(\frac{c\lambda}{\sigma_R}\right)_{i-\frac{1}{2}}^n\frac{E_{r,i}^{n+1}-E_{r,i-1}^{n+1}}{\Delta x}\right\}\nonumber\\
&&-c\sigma_P^n \left[a_r \left\{T_g(\rho_i,e_{g,i}^{n+1})\right\}^4-E_{r,i}^{n+1}\right]+\left\{\mathbb{D}_i^n:(\nabla\mathbf{v})_i\right\}E_{r,i}^{n+1}\Bigr].
\end{eqnarray}
We can rewrite (\ref{nrm}) explicitly:
\begin{eqnarray}
\frac{\partial f_i^g}{\partial e_{g,i}^{n+1}}\delta e_{g,i}^{n+1}+\frac{\partial f_i^g}{\partial E_{r,i}^{n+1}}\delta  E_{r,i}^{n+1}=-f_i^g,\label{nri1}\\
\frac{\partial f_i^r}{\partial e_{g,i}^{n+1}}\delta e_{g,i}^{n+1}+\frac{\partial f_i^r}{\partial E_{r,i}^{n+1}}\delta  E_{r,i}^{n+1}+\frac{\partial f_i^r}{\partial E_{r,i+1}^{n+1}}\delta  E_{r,i+1}^{n+1}+\frac{\partial f_i^r}{\partial E_{r,i-1}^{n+1}}\delta  E_{r,i-1}^{n+1}=-f_i^r.\label{nri2}
\end{eqnarray}
By substituting (\ref{nri1}) into (\ref{nri2}), we can eliminate the equation related to gas energy \citep{zeusmp} (this procedure corresponds to performing partial LU decomposition analytically). Then we obtain the matrix equation:
\begin{eqnarray}
\left(\frac{\partial f_i^r}{\partial E_{r,i}^{n+1}}-\frac{\partial f_i^r}{\partial e_{g,i}^{n+1}}\frac{\partial f_i^g}{\partial E_{r,i}^{n+1}}\Big{/}\frac{\partial f_i^g}{\partial e_{g,i}^{n+1}}\right)\delta  E_{r,i}^{n+1}+\frac{\partial f_i^r}{\partial E_{r,i+1}^{n+1}}\delta  E_{r,i+1}^{n+1}+\frac{\partial f_i^r}{\partial E_{r,i-1}^{n+1}}\delta  E_{r,i-1}^{n+1}
=\frac{\partial f_i^r}{\partial e_{g,i}^{n+1}}\Big{/}\frac{\partial f_i^g}{\partial e_{g,i}^{n+1}}f_i^g-f_i^r.\label{nrif}
\end{eqnarray}
The derivatives are given as follows:
\begin{eqnarray}
\frac{\partial f_i^g}{\partial e_{g,i}^{n+1}}&=&1+4\Delta t\, c\sigma_P^n a_r [T_g(\rho_i,e_{g,i}^{n+1})]^3\frac{\partial T_g}{\partial e_{g}}(\rho_i,e_{g,i}^{n+1}), \\
\frac{\partial f_i^g}{\partial E_{r,i}^{n+1}}&=&-\Delta t\, c \sigma_P^n, \\
\frac{\partial f_i^r}{\partial e_{g,i}^{n+1}}&=&-4\Delta t\, c\sigma_P^n a_r  [T_g(\rho_i,e_{g,i}^{n+1})]^3\frac{\partial T_g}{\partial e_{g}}(\rho_i,e_{g,i}^{n+1}),\\
\frac{\partial f_i^r}{\partial E_{r,i}^{n+1}}&=&1+\Delta t \left[c \sigma_P^n +\mathbb{D}_i^n:(\nabla\mathbf{v})_i -\frac{c}{\Delta x^2}\left\{\left(\frac{\lambda}{\sigma_R}\right)_{i+\frac{1}{2}}^n+\left(\frac{\lambda}{\sigma_R}\right)_{i-\frac{1}{2}}^n\right\}\right],\\
\frac{\partial f_i^r}{\partial E_{r,i\pm 1}^{n+1}}&=&\frac{c\Delta t}{\Delta x^2}\left(\frac{\lambda}{\sigma_R}\right)_{i\pm\frac{1}{2}}^n.
\end{eqnarray}
$T_g(\rho,e_g)$ and $\frac{\partial T_g}{\partial e_g}(\rho,e_g)$ are given from the tabulated EOS. Note that this Jacobi matrix is symmetric. We can obtain the solution by updating $E_{r,i}$ and $e_{g,i}$ using $\delta E_{r,i}$ and $\delta e_{g,i}$ calculated from (\ref{nrif}) and (\ref{nri1}) until $f_i$ and $\delta \mathbf{X}$ become sufficiently small. As the initial guess for $E_{r,i}^{n+1}$ and $e_{g,i}^{n+1}$, we adopt the solution at the timestep $n$. In our simulations, we use convergence thresholds like $\max\left(\frac{f_i^g}{e_{g,i}},\frac{f_i^r}{E_{r,i}}\right)< 5\times 10^{-4}$ and $\max\left(\frac{\delta e_{g,i}}{e_{g,i}},\frac{\delta E_{r,i}}{E_{r,i}}\right)< 5\times 10^{-4}$. If these thresholds are not satisfied after many iterations, we take substeps with shorter $\Delta t$ and try again until we obtain the converged solution successfully.

\subsubsection{Linear System Solver}
In three dimensional Cartesian coordinate, the Jacobi matrix in (\ref{nrif}) is a very large sparse seven-diagonal matrix. Moreover, because of the strong nonlinearity of the system, the matrix is not necessarily diagonally dominant. Therefore we need an efficient and robust sparse matrix solver and extensive computational resources to solve this large (typically $64^3=262144$ cells per grid level) linear system. We found that the combination of the BiCGStab (stabilized bi-conjugate gradient) solver and the incomplete LU (ILU) decomposition preconditioner without fill-in works well for our problems\footnote{ILU type preconditioners are very robust and efficient, significantly reducing the number of iterations required in iterative solvers. However, it does not easily fit parallelization because of the dependencies between the operations. Therefore, although it is a good algorithm for supercomputers with high single node performance like NEC SX-9 which we use, its scalability is problematic for massive parallel architectures.}.

\subsubsection{Radiation Force}
We simply integrate the radiation force terms in Step 5 using the obtained solution in Step 4. These terms are relatively small, at least in the early phase of low-mass star formation processes.

\subsection{Nested-Grid}
In order to achieve the huge dynamic range required in protostellar collapse simulations, we adopt the three dimensional nested-grid technique \citep{ybl93,yk95,zy97,mh03b,mtm04}. This is a simplified version of the adaptive mesh refinement (AMR) technique. Each grid level consists of $N_x\times N_y\times N_z$ cubic cells. The number of the cells in one direction $N_*$ must be a power of 2. The finer grid is placed around the center of the coarser grid self-similarly. The size of a finer cell is half of that of a coarser cell. We number the levels from coarsest to finest: $l=1({\rm coarsest}),2,...,L({\rm finest})$.

We calculate the coarsest grid $l=1$ first and then proceed to finer grid levels because we require the boundary conditions at the next timestep in the implicit update. We update all the grid levels in each step described in \ref{os}. The timestep is determined by the Courant-Friedrich-Levy (CFL) criterion derived for the MHD part. All the grids share this common timestep and are advanced synchronously. We adopt so-called Jeans condition proposed by \citet{trlv97} (see also \citet{com08}) for refinement; we generate a finer grid to resolve the minimum Jeans length with 16 cells.

For the MHD and resistivity parts, we apply the standard procedures to the boundaries between the levels of different resolution. That is, when we update a level $l$, we construct the boundaries from the coarse level $l-1$ using time and spatial interpolation. For the spatial interpolation, we adopt linear interpolation with a slope limiter to assure the monotonicity. After updating all the levels, we transfer the results in the overlapped regions from the finer grid to the coarser grid using conserved variables and recalculate the flux in the coarser level $l-1$ using the obtained flux in the finer level $l$, conserving the total flux across the level boundaries.

\subsubsection{Implicit Update on Nested-Grid Hierarchy}
The implicit time integration for the radiation transport on the nested grids is more tricky. Because all the cells among the whole grid levels interact with each other in a single timestep in the implicit scheme, in principle we should integrate all the levels at the same time treating them like one huge non-uniform grid. However, it makes the Jacobian matrix irregular and complicated, and the computational costs can be very expensive because a huge number of cells are involved in the integration. Therefore we solve each grid separately as we do in the MHD part. This enables us to adopt computational algorithms highly optimized for regular sparse matrices.

When we update a grid level $l$, we treat it as a uniform grid. The boundary conditions at the next timestep are constructed using linear spatial interpolation from the coarser grid which is already updated, and treated as fixed (Dirichlet) boundaries. In (\ref{flds}) and (\ref{fld2}), we need to estimate the internal energy and temperature, rather than the total energy. We use the total energy for grid interaction in the MHD part, but it causes artificial thermalization of the internal dispersion of fluid motion and magnetic fields resulting in unphysically high entropy and temperature in the region overlapped with the finer grids. To avoid this, we take the temperature average of the nearest eight cells in the overlapped finest levels and calculate the internal energy using the gas density and the averaged temperature. For consistency between the gas and radiation, we also take the radiation temperature averaged over the nearest eight cells in the finest level. These procedures violate conservation of the total energy, but in the diffusion approximation it is more important to properly estimate the temperature at the cell center and its gradient, rather than the total energy. And its effects are kept small because we overwrite the overlapped regions with the results obtained in the finer levels at the end of every timestep, which satisfy the local conservation laws.

We only consider the interaction between the coarse and fine levels at the level boundaries in our implicit scheme on the nested grids, but it may cause the loss of consistency and accuracy. In our production runs, we confirmed that this treatment only causes minor discrepancies between the grid levels in the extremely optically-thin regions. Because we are mainly interested in the evolution of the condensed objects, we can tolerate these errors.

\subsection{Boundary Conditions}
As the outer-most boundary conditions for the magnetohydrodynamics and radiation parts, we set all the cells outside the initial Bonnor-Ebert sphere to maintain their initial values, mimicking an isolated molecular cloud core confined in a static environment. For the Poisson's equation of self-gravity, we compute the gravitational potential of the isolated system at the boundaries by the multi-pole expansion \citep{mh03}. Our boundary conditions allow the gas to inflow into the computational domain through the boundaries, but the inflowing mass during the simulation is sufficiently smaller than that of the total mass of the initial cloud, about 8\% in all the models.

\section{Simulation Setups}
We use unstable Bonnor-Ebert \citep[][hereafter BE]{bonnor,ebert} spheres of $T=10\, {\rm K}$as the initial conditions of our simulations, mimicking isolated molecular cloud cores. We construct a critical BE sphere (with a dimension less radius of $\xi=6.45$) and make it unstable by increasing the gas density by a factor of $A_0$. Then we introduce uniform (rigid-body) rotation, uniform magnetic fields and $m=2$ density perturbation, where $m$ is the number of longitudinal modes. The initial density profile is given as follows:
\begin{eqnarray}
\rho(r)=\rho_c\rho_0(r)(1+A_0)\left[1+A_2\frac{r^2}{R^2}\cos(2\phi)\right],
\end{eqnarray}
where $\rho_c$ is the gas density at the center of the cloud core, $\rho_0(r)$ the normalized density profile of the critical BE sphere, $R$ the radius of the critical BE sphere, $A_2$ the amplitude of $m=2$ perturbation, respectively. In order to minimize the effect of initial resolution, we adopt this ``regularized" $m=2$ perturbation which is smooth at the center of the cloud in contrast to \citet{bb79}. 

In this work, we adopt $\rho_c=10^{-18}\,{\rm g \, cm^{-3}}$ and $A_0=0.2$, which yield an unstable BE-like sphere whose mass and radius are $M\sim 1M_\odot$ and $R\sim 4.25\times10^{-2}\,{\rm pc}\sim 8800\,{\rm AU}$. The initial free fall time at the center of the cloud is $t_{\rm ff}\sim 6.08\times 10^4 \,{\rm yrs}$.

We calculated a spherically symmetric model and four magnetized rotating models with and without Ohmic dissipation; the parameters of the models are summarized in Table~\ref{models}. The first letter of the model name denotes the treatment of magnetic fields: {\it I} denotes an ideal MHD model and {\it R} a resistive MHD model. The second letter represents the initial rotation speed: {\it F} is fast and {\it S} is slow. Note that we choose these parameters so that the first core disks do not fragment before the second collapse because of the limitation of our nested-grids. Model {\it SP} is the spherical model without rotation and magnetic fields. For magnetized models, we impose the uniform magnetic fields of $20\,{\rm \mu G}$ parallel to the rotation axis. The corresponding mass-to-flux ratio normalized by the critical value of stability is $\mu_0\equiv \frac{M/\Phi}{(M/\Phi)_{\rm crit}}\sim 3.8$ where $\Phi=\pi R^2 B_0$ and $(M/\Phi)_{\rm crit}=\frac{0.53}{3\pi}\left(\frac{5}{G}\right)^{1/2}$. Here we adopt the critical mass-to-flux ratio of \citet{ms76} but we should regard this value as just a guide because our initial conditions are not uniform. There is another similar threshold for stability between gravity and magnetization derived for disks \citep{nn78}; we have $\lambda_0\equiv\frac{\Sigma_0/B_0}{(\Sigma/B)_{\rm crit}}=7.2$ at the center of the cloud, where $(\Sigma/B)_{\rm crit}=(4\pi^2 G)^{-1/2}$. These mass-to-flux ratios indicate that our magnetized models are in the magnetically super-critical regime but considerably magnetized.

We define the origin of time as the epoch of formation of the protostellar core, i.e., when the central gas density exceeds $\rho_c = 10^{-3} \, {\rm g\, cm^{-3}}$ for the first time, for descriptive purpose. We stop our simulations when the central temperature reaches $T_c \sim 10^5\,{\rm K}$. Our simulations show the formation and earliest evolution of protostars. The typical spatial resolution around the surface of the first cores ($r \sim 3-5\, {\rm AU}$) is $\Delta x \sim \, 0.14 {\rm AU} \, (L=12)$ and the finest resolution at the end of the simulations is $\Delta x \sim 6.6 \times 10^{-5} \, {\rm AU} \sim 0.014 R_\odot \, (L=23)$. The nested-grid technique enables us to achieve such a huge spatial dynamic range of more than eight orders of magnitude.

\begin{table}[tbp] 
\begin{center}
\begin{tabular}{c|cccccccc}
Model & $\Omega t_{\rm ff}$ & $\Omega \, (\times 10^{-14}\, {\rm s^{-1}})$ & $B_0 \, ({\rm\mu G})$ & $\mu_0$ & $\lambda_0$ & $A_2$ & Resistive?\\
\hline
{\it SP} & 0 & 0 & 0 & $\infty$ & $\infty$ & 0 & --\\
{\it IS} & 0.023 & 1.2 & 20 & 3.8 & 7.2 & 0.1 & N\\
{\it IF} & 0.046 & 2.4 & 20 & 3.8 & 7.2 & 0.1 & N\\
{\it RS} & 0.023 & 1.2 & 20 & 3.8 & 7.2 & 0.1 & Y\\
{\it RF} & 0.046 & 2.4 & 20 & 3.8 & 7.2 & 0.1 & Y
\end{tabular}
\end{center}
\caption{Summary of the initial model parameters. From left to right: the normalized angular velocity, the angular velocity, the magnetic field strength, the averaged mass-to-flux ratio, the local mass-to-flux ratio at the cloud center, the amplitude of $m=2$ perturbation and whether the resistivity is introduced or not. Other parameters are common: $M=1\,M_\odot$, $R\sim 8800\, {\rm AU}$, $\rho_c=1.2\times 10^{-18}\, {\rm g\, cm^{-3}},$ and $T_0=10\,{\rm K}$. See the text for details.}
\label{models}
\end{table}

\section{Results}
\subsection{Spherical Model}
We first show the results of the spherical model {\it SP} to understand the whole evolution from a molecular cloud core to a protostellar core, and to demonstrate validity of our code. 

\begin{figure}[tb]
\begin{center}
\scalebox{0.32}{\includegraphics{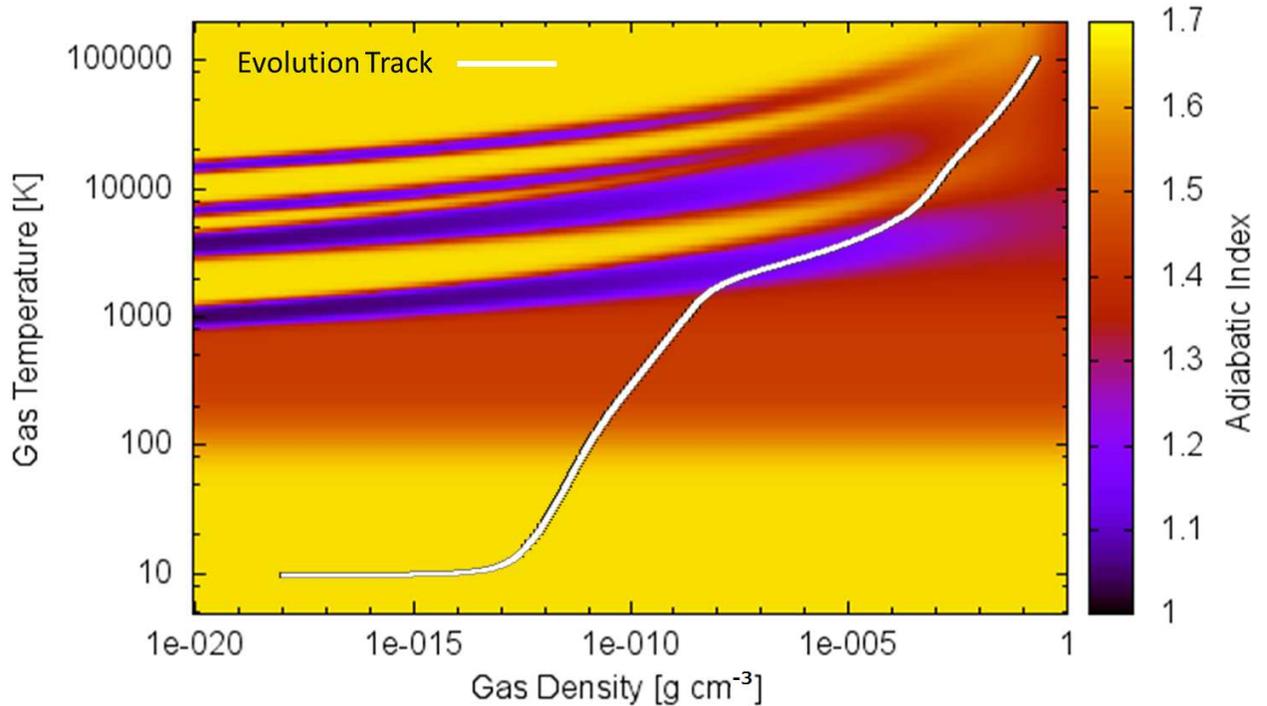}}
\caption{The evolution track of the central gas element (white) in {\it SP} overplotted on the distribution of the adiabatic index $\Gamma$ in the $\rho-T$ plane. Four low-$\Gamma$ (blue) bands correspond to the endothermic reactions of the dissociation of molecular hydrogen, the ionization of hydrogen, the first and second ionization of helium, from bottom to top.}
\label{s_ee}
\end{center}
\end{figure}

\begin{figure}[htb]
\begin{center}
\scalebox{1.25}{\includegraphics{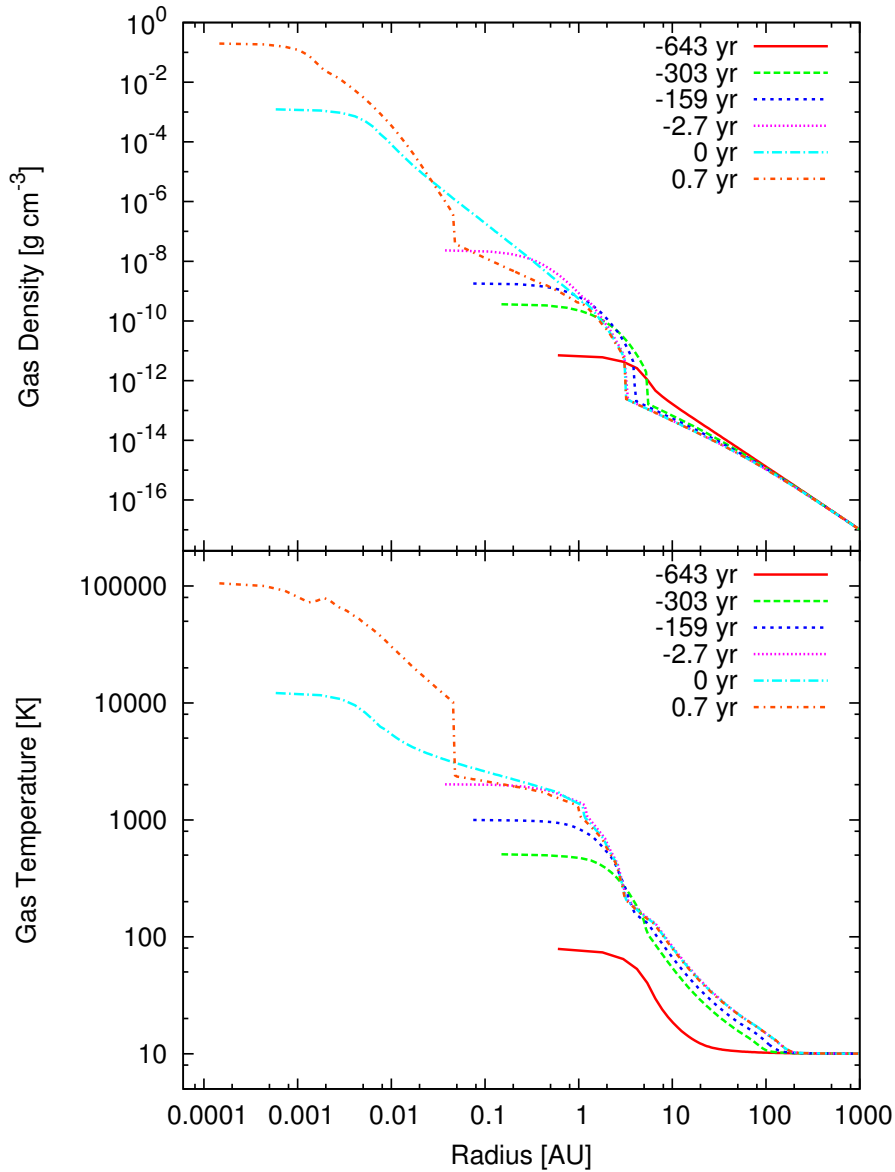}}
\caption{The evolution of the gas density (top) and temperature (bottom) profiles in Model {\it SP}. The numbers show the order of evolution.}
\label{s_dt}
\end{center}
\end{figure}

\begin{figure}[htb]
\begin{center}
\scalebox{1.5}{\includegraphics{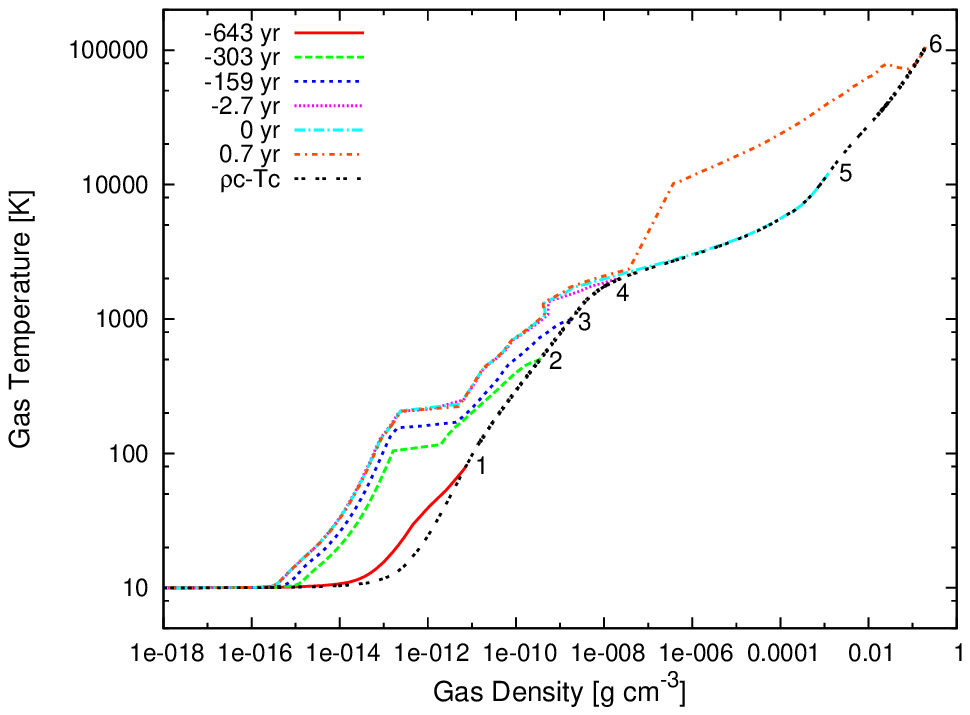}}
\caption{The distributions of the thermodynamic properties in the $\rho-T$ plane. The evolution track of the central gas element is also plotted.}
\label{s_rhot}
\end{center}
\end{figure}

\subsubsection{Thermal Evolution and EOS}

Before discussing the global structure and evolution, we explain the thermal evolution of the central gas element. To see the effects of the microphysics involved in the EOS on the thermal evolution, we show the evolution track of the central gas element in the $\rho-T$ plane and the adiabatic index $\Gamma$ in Figure~\ref{s_ee}. The evolution is consistent with the results of the more elaborate 1D spherically-symmetric RHD simulations \citep{mi00} except for the details of the EOS. From this figure, we can understand the relation between the thermal evolution of the gas and the microphysical processes.

While the gas density is low ($\rho_c \ltsim 10^{-13}\, {\rm g\, cm^{-3}}$), the gas collapses isothermally because radiation cooling is very efficient, and the EOS does not matter in this regime. When the gas density gets sufficiently high and radiation cooling becomes inefficient, the gas evolves quasi-adiabatically and the temperature rises. Beyond this point, the EOS plays almost dominant roles in the thermal evolution of the gas (at the center of the cloud; note that the gas in outer region is affected by radiation transfer). While the gas is still cold ($T_c \ltsim 100\, {\rm K}$), the adiabatic index $\Gamma$ is about 5/3 because the rotational transitions cannot be excited and molecular hydrogen behaves like monoatomic molecules. $\Gamma$ decreases to $\sim 7/5$ when the gas becomes warm enough to excite rotation. When the temperature exceeds $T_c\sim 2000\, {\rm K}$, molecular hydrogen starts to dissociate and the second collapse begins. The endothermic reaction of hydrogen molecule dissociation significantly decreases the effective adiabatic index to $\Gamma \sim 1.1$, below the critical value of stability of a self-gravitational sphere, $\Gamma_{\rm crit}=4/3$. After the completion of the dissociation, the gas evolves quasi-adiabatically again, while the ionizations of hydrogen and helium slightly affect the evolution making the adiabatic index softer. From Figure~\ref{s_ee}, we can see that the ionizations of hydrogen and helium do not cause the ``third collapse"; those endothermic reactions proceed gradually because the reverse reactions (recombinations) occur rapidly in the high density regions and the adiabatic index remains larger than the critical value.

In this work we assume the ortho:para ratio of molecular hydrogen to be 3:1, but if we adopt the equilibrium ratio, additional energy is consumed to convert parahydrogen to orthohydrogen ($\Delta E / k \sim 170\,{\rm K}$), resulting in the softer adiabatic index \citep{boley07,stm09}. The gas does not forget the thermal history in this phase because its evolution is almost adiabatic, therefore the ortho-para ratio has non-negligible impact on the whole star formation process. We have to keep it in mind that the differences in the EOS quantitatively affect the evolution and properties of the first and protostellar cores such as their radii, masses and stability against fragmentation.

\subsubsection{First Core}

The first core is formed at $t\sim -650\,{\rm yr}$ when the central density exceeds $\rho_c \sim 10^{-12}\,{\rm g \, cm^{-3}}$. Its radius is initially about $\sim 5.4\,{\rm AU}$, and it contracts gradually to $\sim 3\,{\rm AU}$ as it evolves (Figure~\ref{s_dt}). The evolution and structure of the first core are in good agreement with the results of the 1D spherically symmetric RHD simulations using full non-gray radiation transfer \citep{mmi98,mi00} and the 3D SPH RHD simulations using the gray FLD approximation \citep{wb06}. The outer region attains a higher entropy than the central gas element because of heating at the shock and radiation from the central hot region (Figure~\ref{s_rhot}). The shock at the surface of the first core is isothermal, i.e., it is a super-critical shock at which almost all the kinetic energy is radiated away to the upstream, as discussed in \citet{com11a}. The lifetime of the spherical first core is a bit longer than that in \citet{bate11}, probably because of the different initial conditions; \citet{bate11} adopted a uniform sphere with higher central gas density, which is more unstable and gives higher accretion rate than our BE sphere.

When the gas temperature exceeds the evaporation temperature of all the dust components ($T\sim 1500\,{\rm K}$), the opacities around the central region drop significantly. The temperature distribution within this dust free region becomes almost flat. As Figure~\ref{s_dt} indicates, the dust evaporation front is located at $R\sim 1.2\,{\rm AU}$ at the end of the first core phase ($t\sim -2.7\,{\rm yr}$). The dust evaporation front is also visible as a small jump in Figure~\ref{s_rhot}. \citet{sch11} pointed out the importance of the dust evaporation on the dynamics of first cores, but it is not prominent in the spherically symmetric model.

\subsubsection{Protostellar Core}
The second collapse begins when the central temperature exceeds $T_c\sim 2000\,{\rm K}$. Soon after the onset of the second collapse, the protostellar core is formed in the short dynamical timescale of several years (the free fall time corresponding to the central density when the second collapse starts ($\rho_c\sim 10^{-8}\,{\rm g\, cm^{-3}}$) is only about 0.67 years). Within 0.7 years after the formation of the protostellar core, it acquires $M_{\rm PC}\sim 2 \times 10^{-2}\,M_\odot$ and the averaged accretion rate is very high, $2.7 \times 10^{-2} M_\odot \,{\rm yr^{-1}}$. The protostellar core expands due to the addition of newly accreted gas and the accretion shock at the surface of the protostellar core quickly propagates outward. The jump condition at this shock is almost adiabatic, which means that the flow is radiatively inefficient (``hot accretion") in this early phase. The outer region of the protostellar core is heated up by the shock and attain a high entropy, leading the protostellar core to be convectively stable \citep{sst1,sst2}. These thermal properties and expansion of protostellar cores cannot be reproduced by the barotropic approximation in which the shock heating is not taken into account. At the end of the simulation, the radius of the protostellar core is about $R_{\rm PC} \sim 0.047\,{\rm AU} \sim 10 R_\odot$. From the virial theorem, the energy released in this phase can be estimated to be $\frac{GM_{\rm PC}^2}{2R_{PC}}\sim 6.8\times 10^{43} \,{\rm erg}$, which is consistent with the total dissociation energy of molecular hydrogen, $\frac{XM_{PC}}{2m_{\rm H}}\chi_{\rm dis}\sim 5.7\times 10^{43}\, {\rm erg}$ where $\chi_{\rm dis}=7.17\times 10^{-12} \, {\rm erg}$ is the dissociation energy of ${\rm H_2}$ \citep{liu09}.

This radius at the end of the simulation is about 2.5 times larger than the radius of the protostellar core obtained in \citet{mi00} ($\sim 4R_\odot$), and still increasing. The expansion in the adiabatic accretion phase had been already reported in \citet{lrs69} and also discussed in \citet{sps86} (see also \citet{wn80b}), and our radius is consistent with their results. It continues to expand until almost all the gas in the first core has accreted onto the protostellar core when the accretion rate gets significantly low and the optical depth of the envelope becomes low enough for radiation cooling. It will take about the average free-fall time of the whole first core, $t_{\rm ff,FC}=\sqrt{\frac{3\pi}{32G \rho_{\rm FC}}}\sim 5.5\, {\rm yrs}$ where $\rho_{\rm FC}=\frac{3M_{\rm FC}}{4\pi R_{\rm FC}^3}=1.5\times 10^{-10}\,{\rm g\, cm^{-3}}$, $M_{\rm FC}\sim 3 \times 10^{-2} M_\odot$ and $R_{\rm FC}\sim 3\,{\rm AU}$. Therefore this expansion is a highly transient phenomenon. In other words, our protostellar core has not settled yet. Note that the gas does not flow outward in this expansion phase, but the newly accreted gas is loaded on top of the protostellar core. It is different from the violent explosion, or ``hiccup" of protostellar cores discussed years ago \citep[e.g.,][]{boss89}.

\subsection{Rotating Models}
\subsubsection{Overview}
\begin{figure}[tb]
\begin{center}
\scalebox{1.25}{\includegraphics{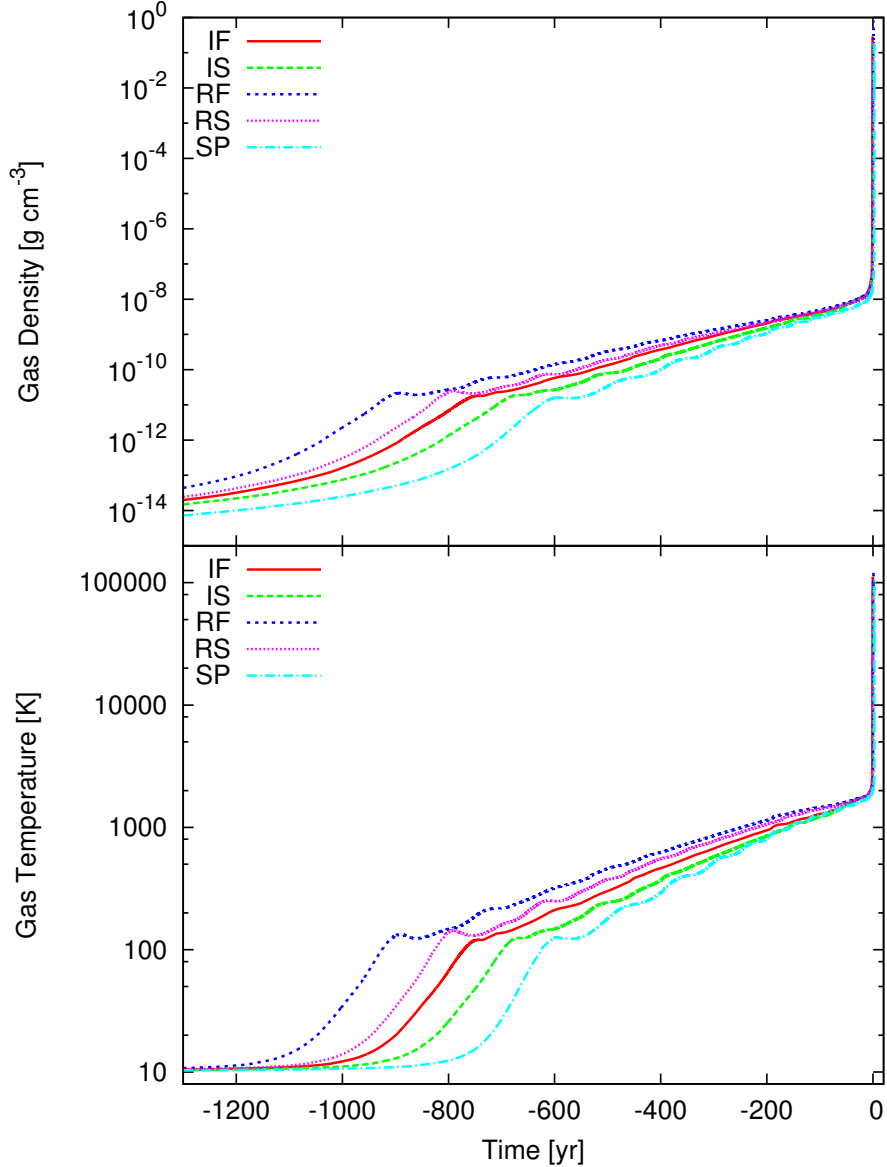}}
\caption{The evolution of the central gas density (top) and temperature (bottom) as functions of time.}
\label{rot_rt}
\end{center}
\end{figure}

\begin{figure}[htb]
\begin{center}
\scalebox{1.}{\includegraphics{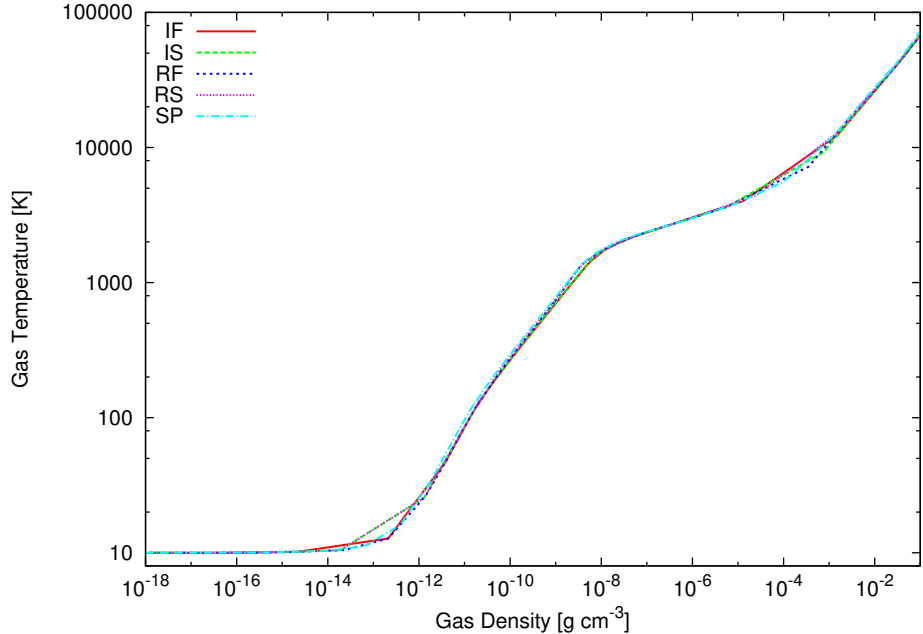}}
\caption{The evolution tracks of the central gas elements in the $\rho-T$ plane.}
\label{rot_rhot}
\end{center}
\end{figure}

We show the evolution tracks of the central gas density and temperature as functions of time in Figure~\ref{rot_rt}, and the evolution tracks in the $\rho-T$ plane in Figure~\ref{rot_rhot}. The first cores are formed when the central density exceeds $\rho_c \gtsim 10^{-12}\,{\rm g\, cm^{-3}}$ and the second collapse begins when $\rho_c \gtsim 10^{-8}\,{\rm g\, cm^{-3}}$. The lifetimes of the first cores are about 650, 720, 800, 850 and 950 years in {\it SP, IS, IF, RS} and {\it RF}, respectively. The presence of rotation extends the first core lifetime but its effect is not significant compared to non-magnetized cases \citep{saigo08,bate11,tomida10b}. The resistive models have slightly longer lifetimes because magnetic fields are weakened by Ohmic dissipation and the efficiency of the angular momentum transport is reduced, but they are still not very long-lived.

Interestingly, all the models show essentially the same thermal evolution (Figure~\ref{rot_rhot}). This is because the central regions of the first cores contract almost spherically because of the efficient angular momentum transport via magnetic fields and their evolution converges on the Larson-Penston self-similar solution \citep{lrs69,pen69}. In the central regions of the collapsing cloud cores, the thermal energy is dominantly larger than the magnetic energy and dissipation of magnetic fields (Joule heating) does not affect the thermal evolution significantly. Hence the thermal properties of the formed protostellar cores such as the gas entropy have already been set in the first core phase to some extent and seem to depend weakly on some initial cloud parameters such as rotation and magnetic fields when there present efficient mechanisms of angular momentum transport. Without magnetic fields, the initial conditions have larger impact on the thermal evolutions, particularly when fragmentation occurs due to fast initial rotation \citep{bate11}. The thermal evolution may also depend on many other factors such as initial temperature, dust opacities and mass of the natal molecular cloud core \citep{tomida10b}.

\subsubsection{Outflows and First Cores}
\begin{figure}[p]
\begin{center}
\scalebox{0.4}{\includegraphics{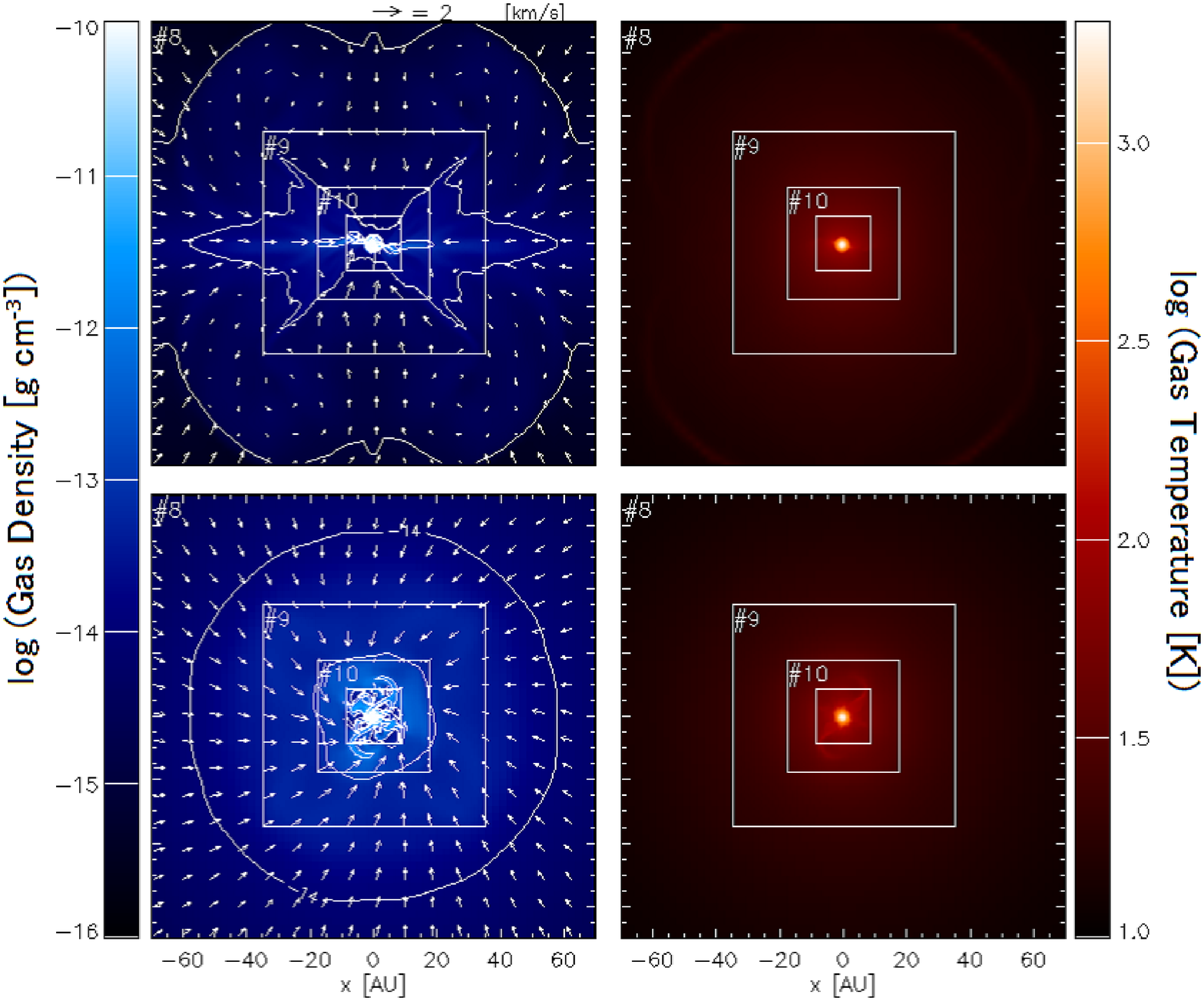}}
\caption{The vertical (top) and horizontal (bottom) cross sections of the gas density (left) and temperature (right) in the outflow scale ($l=8$ or $\sim 140\,{\rm AU}$) of Model {\it IF} just before the second collapse. Projected velocity vectors are overplotted.}
\label{ifof}
\vspace{2em}
\scalebox{0.4}{\includegraphics{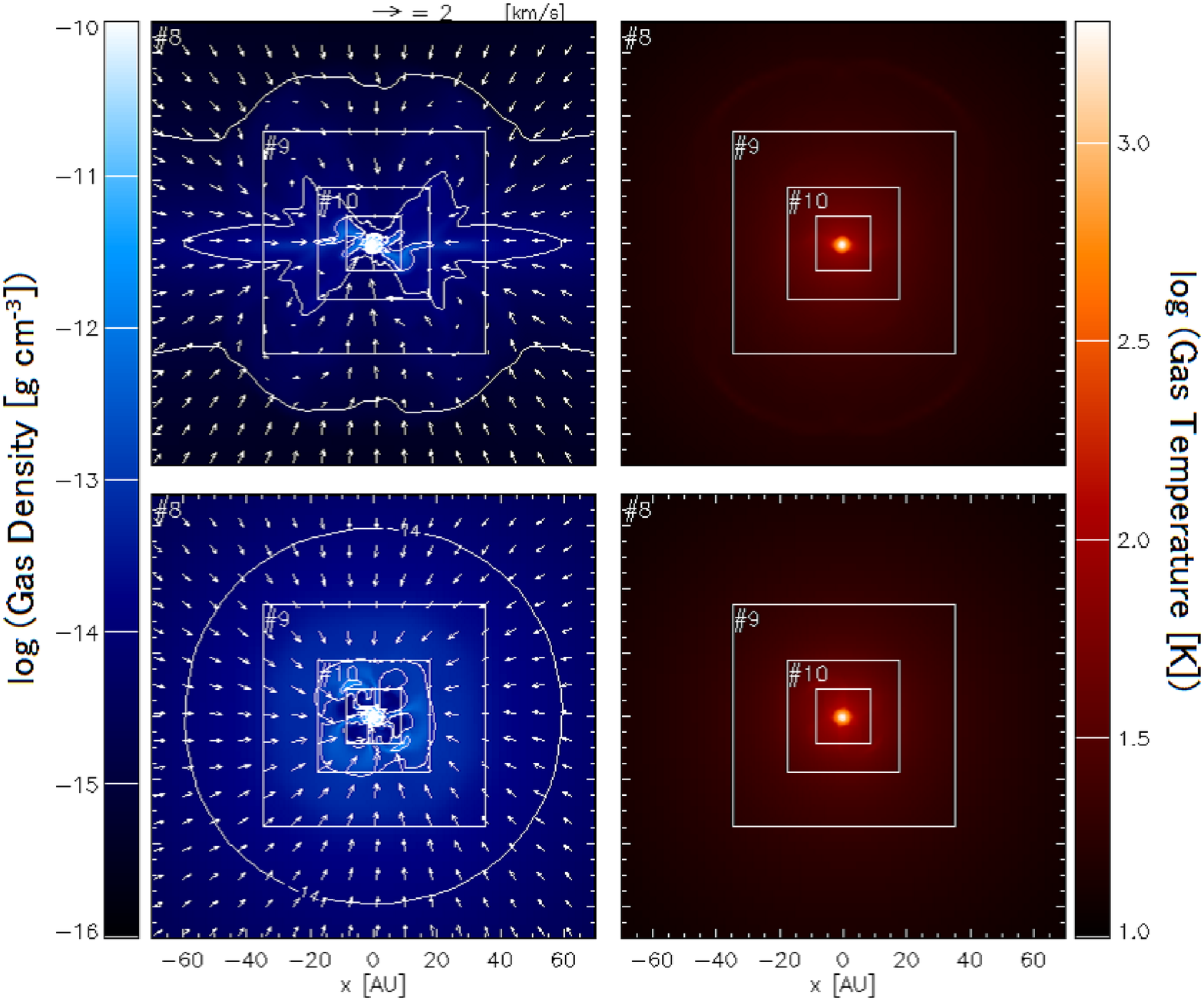}}
\caption{The same as Figure~\ref{ifof} but of {\it IS}.}
\label{isof}
\end{center}
\end{figure}

\begin{figure}[tb]
\begin{center}
\scalebox{0.4}{\includegraphics{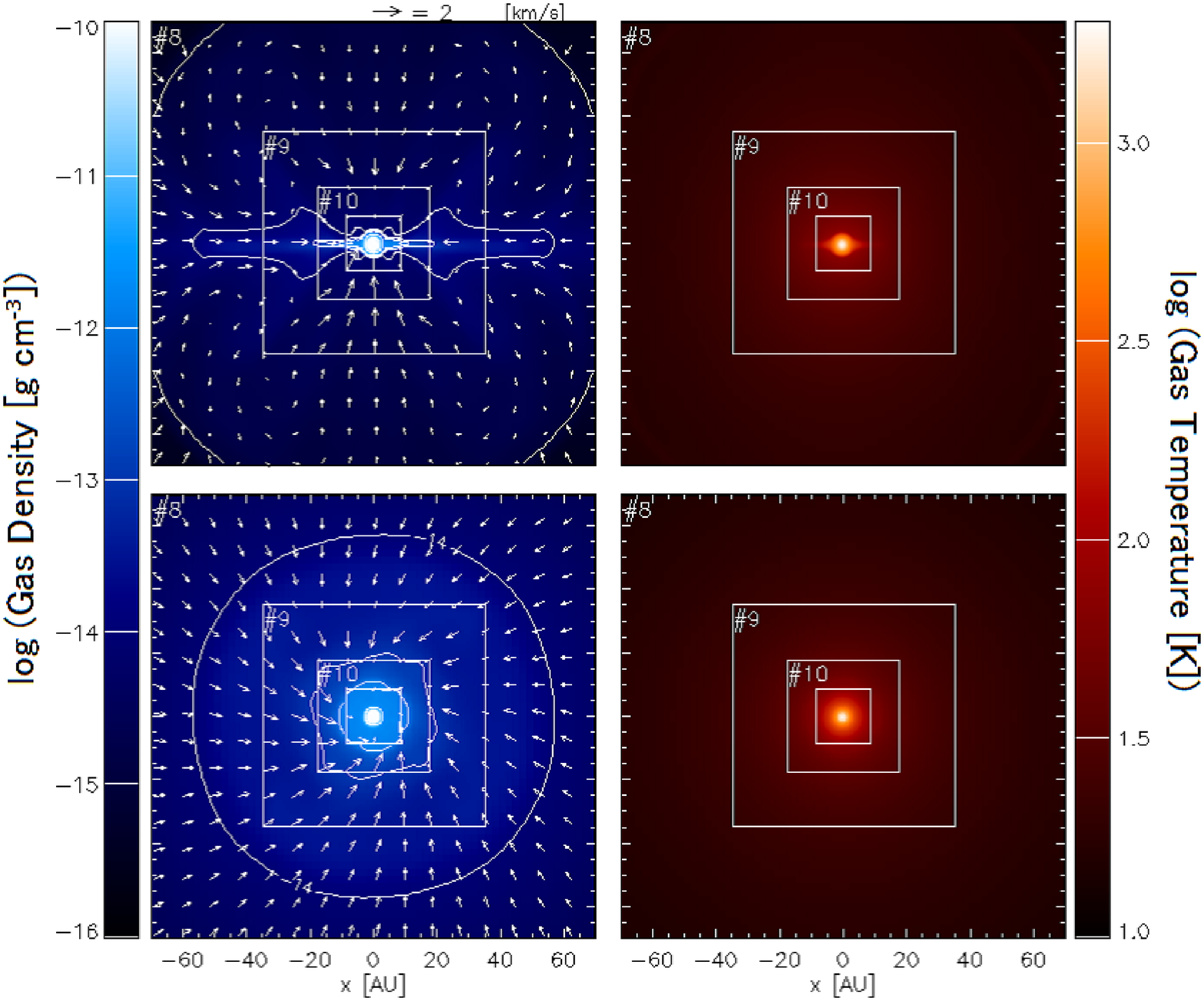}}
\caption{The same as Figure~\ref{ifof} but of {\it RF}.}
\label{rfof}
\vspace{2em}
\scalebox{0.4}{\includegraphics{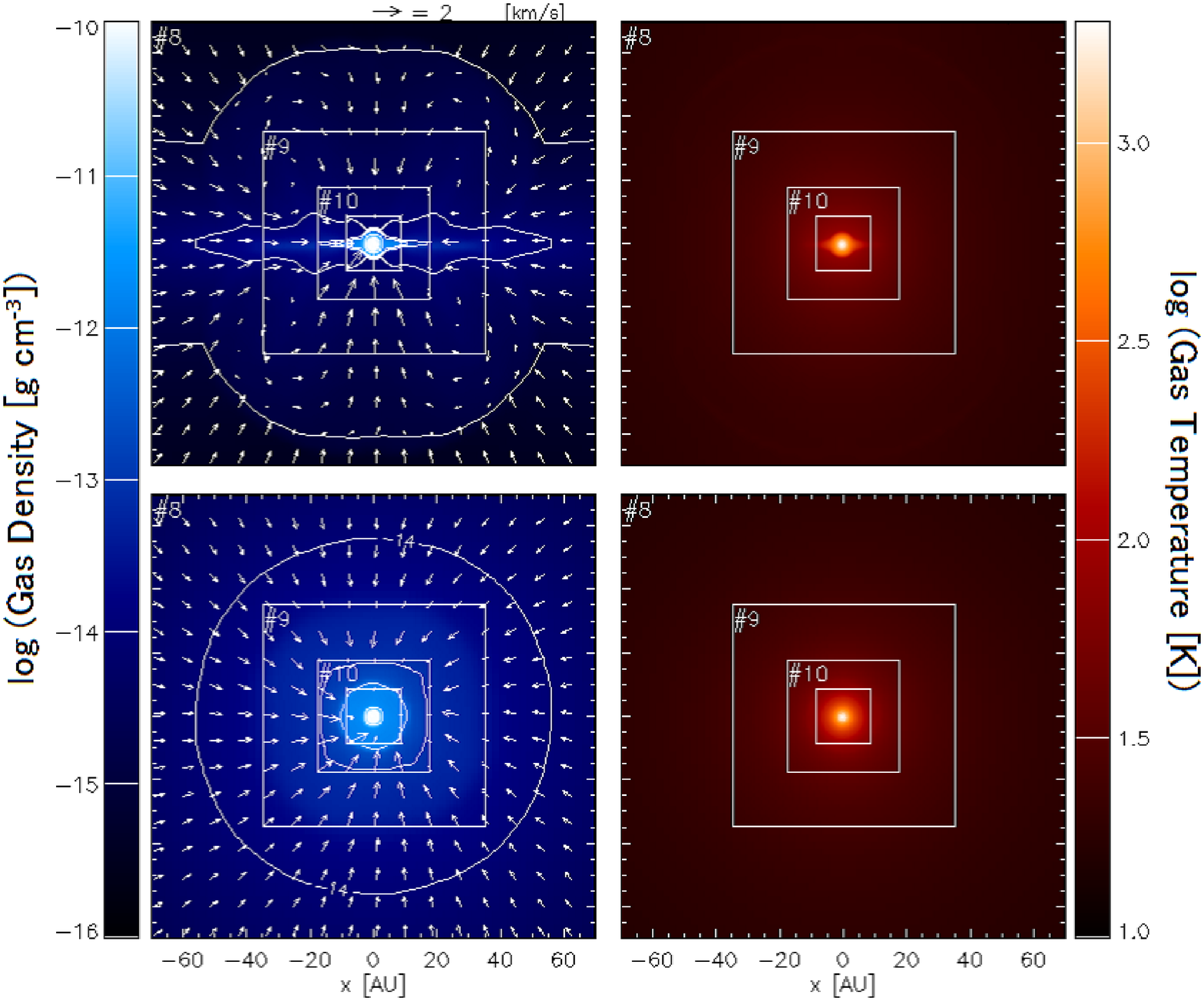}}
\caption{The same as Figure~\ref{rfof} but of {\it RS}.}
\label{rsof}
\end{center}
\end{figure}

\begin{figure}[p]
\begin{center}
\scalebox{0.4}{\includegraphics{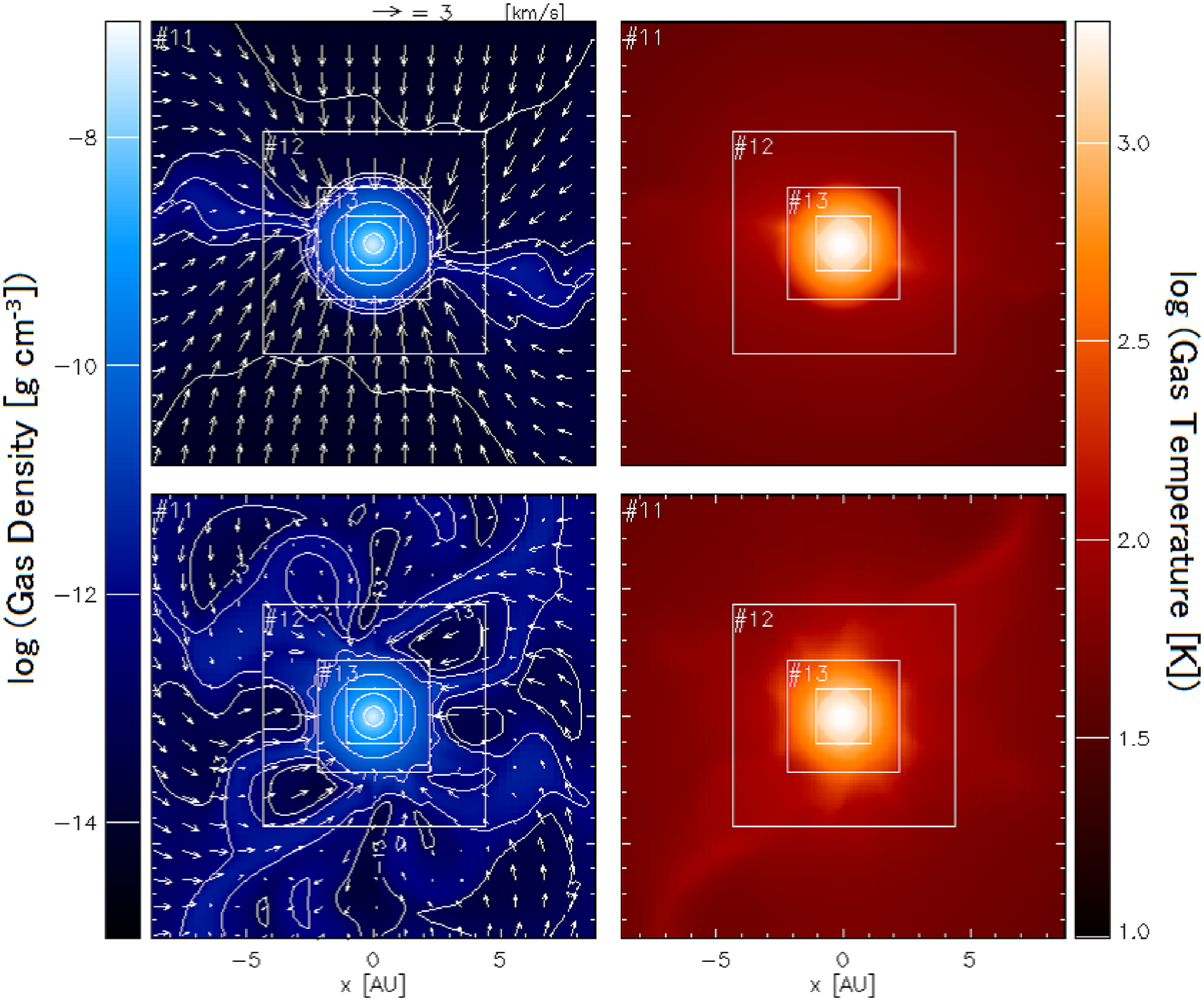}}
\caption{The vertical (top) and horizontal (bottom) cross sections of the gas density (left) and temperature (right) in the first core scale ($l=11$ or $\sim 18\,{\rm AU}$) of Model {\it IF} just before the second collapse.}
\label{iffc}
\vspace{2em}
\scalebox{0.4}{\includegraphics{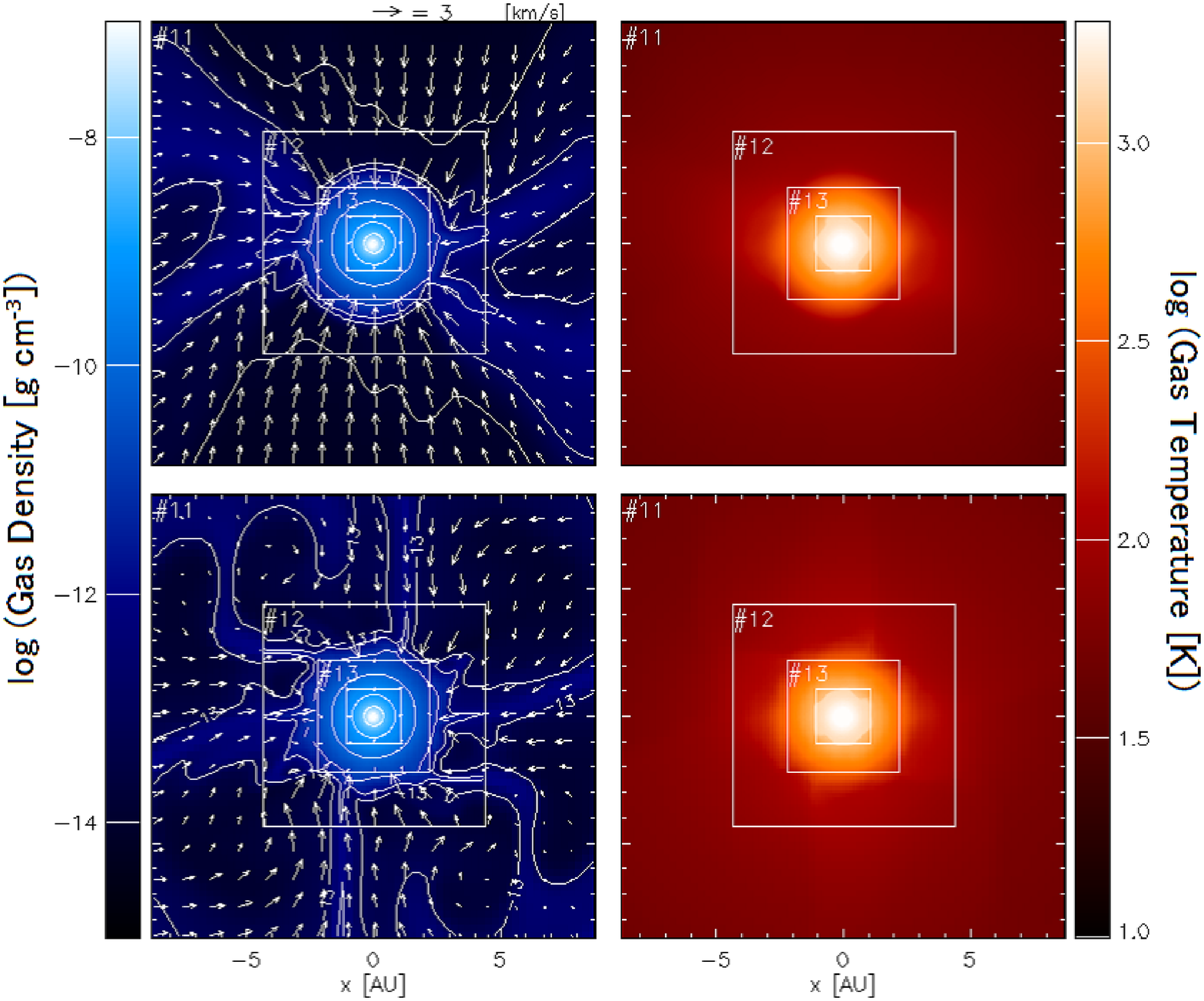}}
\caption{The same as Figure~\ref{iffc} but of {\it IS}.}
\label{isfc}
\end{center}
\end{figure}

\begin{figure}[htp]
\begin{center}
\scalebox{0.4}{\includegraphics{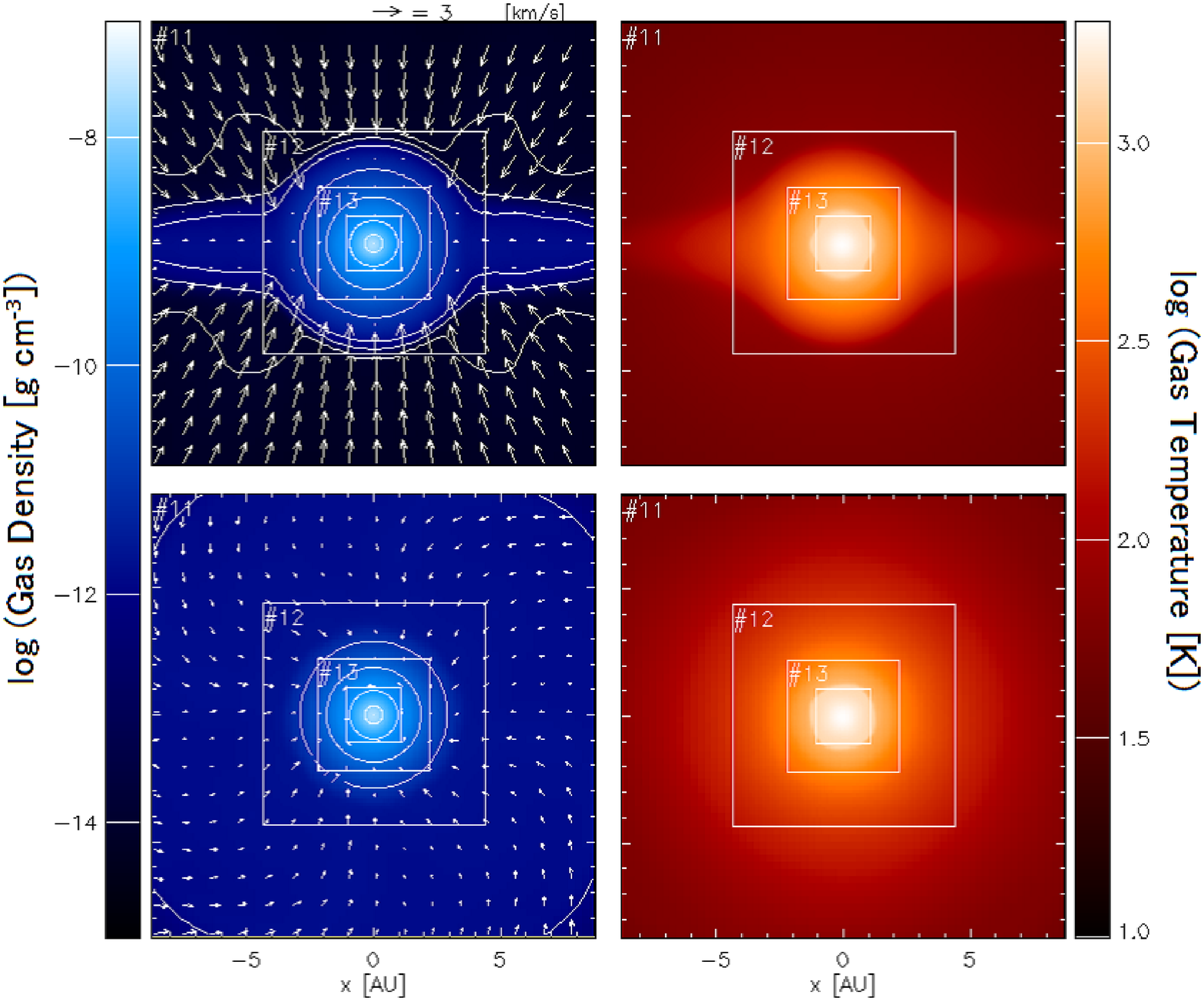}}
\caption{The same as Figure~\ref{iffc} but of {\it RF}.}
\label{rffc}
\vspace{2em}
\scalebox{0.4}{\includegraphics{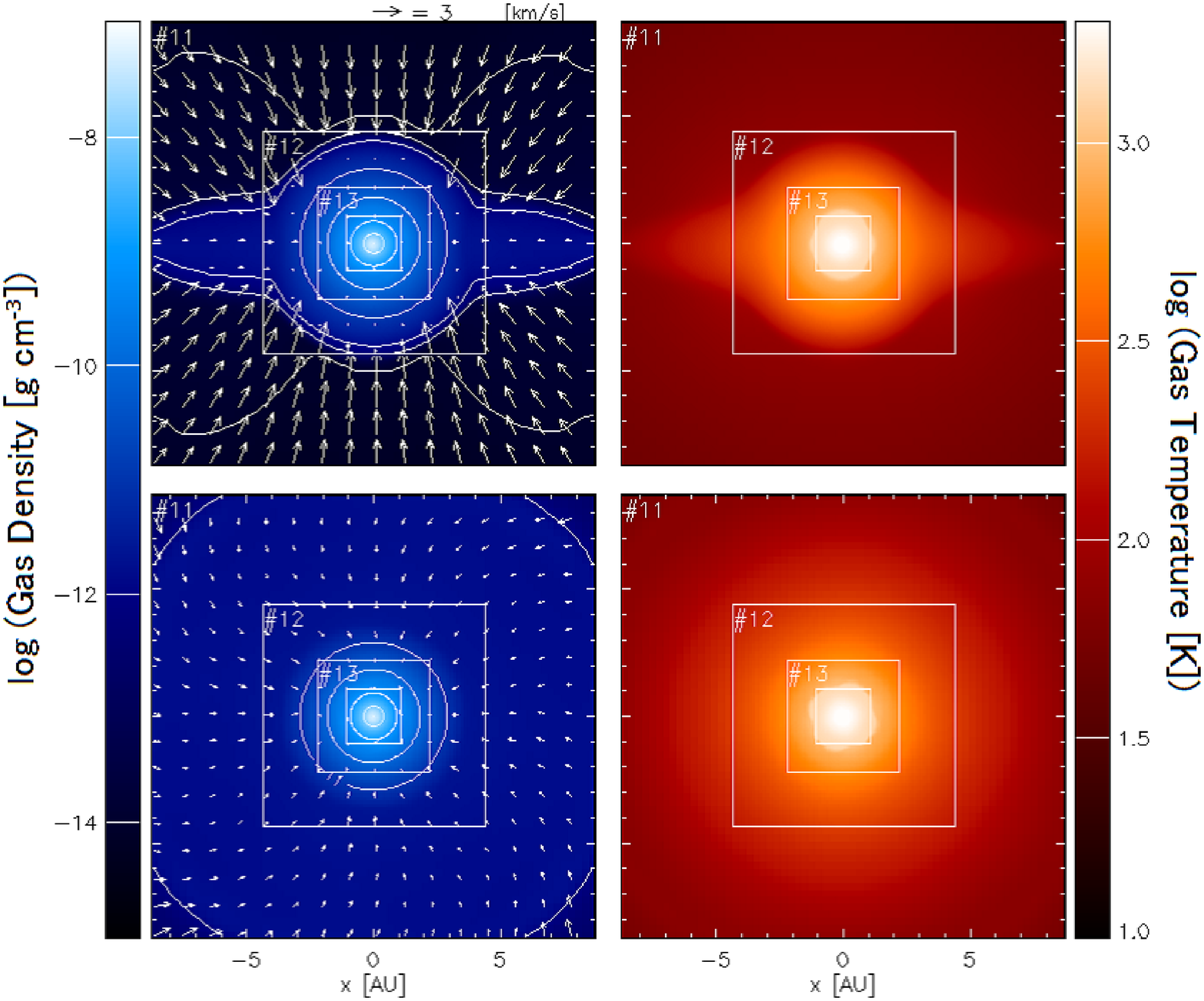}}
\caption{The same as Figure~\ref{rffc} but of {\it RS}.}
\label{rsfc}
\end{center}
\end{figure}

\begin{figure}[p]
\begin{center}
\scalebox{0.6}{\includegraphics{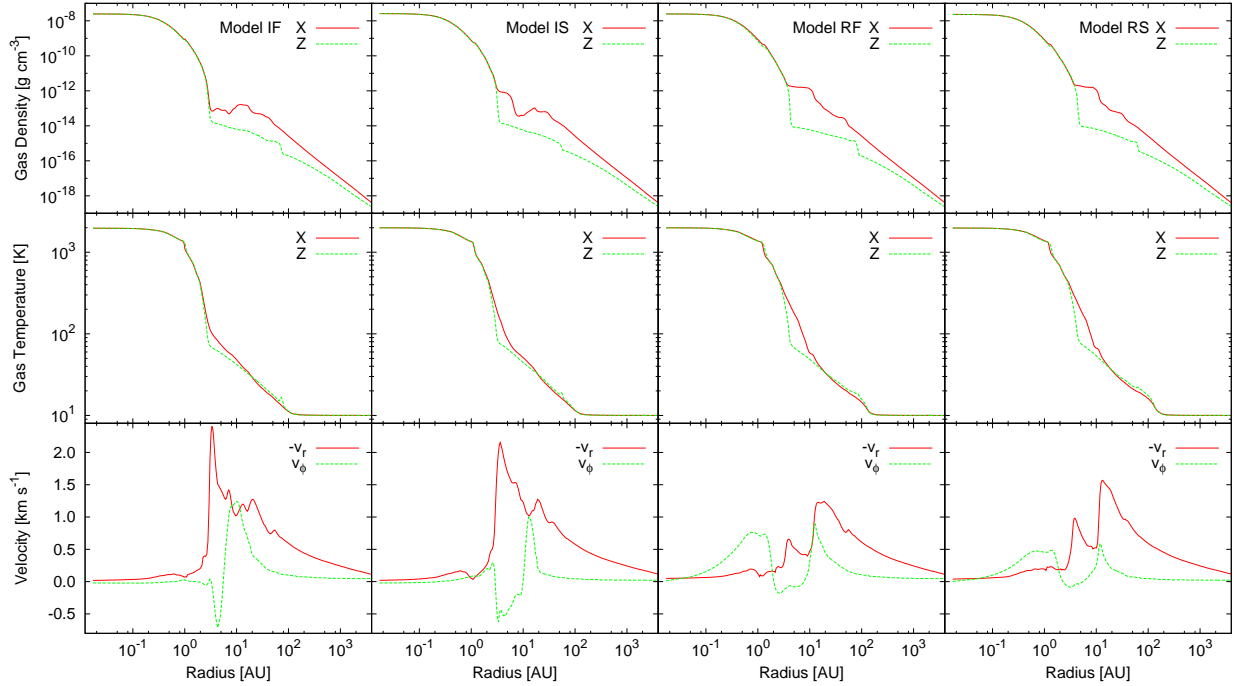}}
\caption{The radial profiles of the gas density, temperature along the $x$- (in the disk mid-plane, red) and $z$-axes (along the rotational axis, green), and the infall (red) and rotation (green) velocities along the $x$-axis (from top to bottom) at the end of the first core phase. From left to right, the different columns are for Model {\it IF}, {\it IS}, {\it RF} and {\it RS}.}
\label{fcprof}
\end{center}
\end{figure}

After the formation of the first cores, slow and loosely-collimated outflows are launched from the first cores by magneto-centrifugal force \citep{bp82,kms98}. We show the density and temperature distributions of the outflow scale (corresponding to the grid level $l=8$ or $\sim 140\,{\rm AU}$) and of the first core scale ($l=11$ or $\sim 18\,{\rm AU}$) in Figures~\ref{ifof} -- \ref{rsfc} at the end of the first core phase. The outflows can be seen as slowly outgoing gas denser than the envelope that extends to $z=55, 70, 60$ and $80 \, {\rm AU}$ in {\it IS}, {\it IF}, {\it RS} and {\it RF}, respectively. The outflows are not prominent in the temperature plots except for the thin shock-heated layers because they are optically thin and radiation cooling/heating is efficient. We also show the profiles of the gas density, temperature and velocities along $x$- and $z$-axes in Figure~\ref{fcprof}. Roughly speaking, the properties of the first cores and outflows are consistent with those in previous studies \citep[][etc.]{com10,tomida10a}.

The properties of the outflows such as velocities and traveling distances are similar in all the models. This is because the driving radii are similar in all the models, $\sim 10 \, {\rm AU}$, where the gas density is not high enough for the resistivity to work. The outflow velocities are comparable to the rotational velocities at this radius, $\sim 1\,{\rm km\, s^{-1}}$. Therefore the traveling distances are almost proportional to the lifetimes of the first cores. It seems difficult to find the effect of the resistivity on the outflow properties, because there is no essential difference in the outflows between the resistive and ideal MHD models \citep[see also][]{yamada09}.

\begin{figure}[htb]
\begin{center}
\scalebox{1.25}{\includegraphics{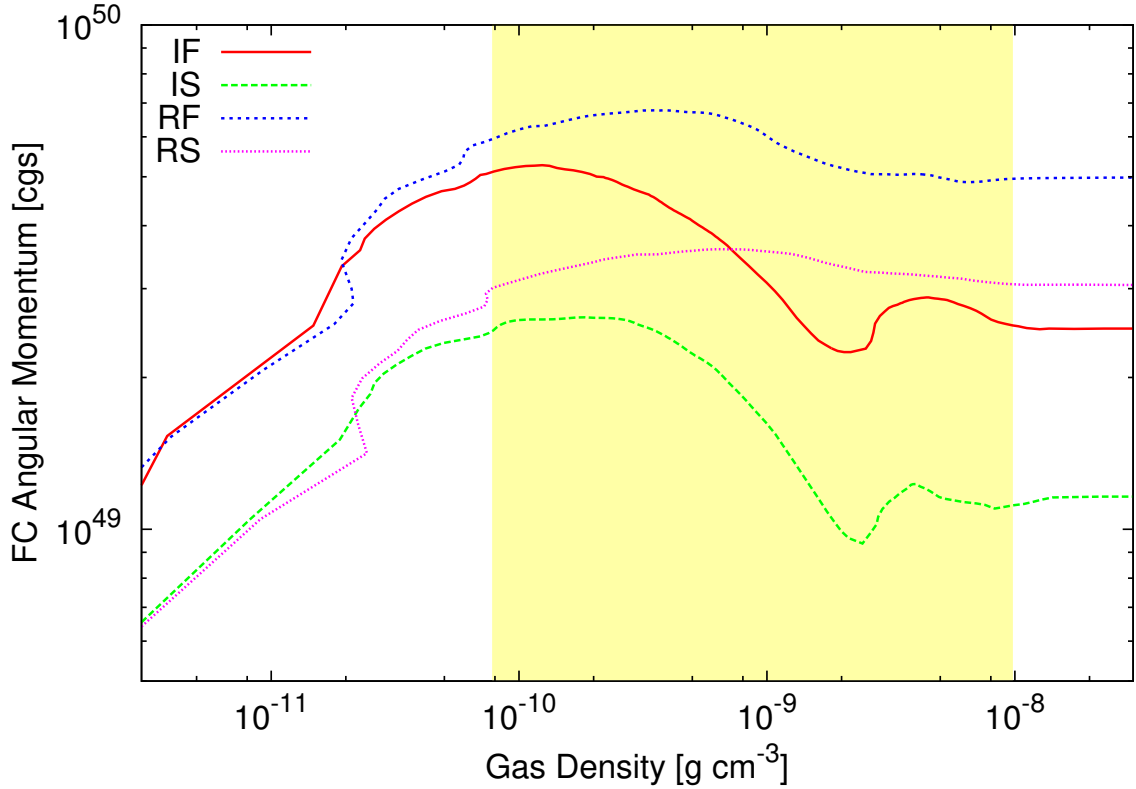}}
\caption{The evolution of the total angular momenta in the first cores versus the central gas density. Magnetic fields are decoupled from fluid where $R_m < 1$ (yellow shaded region).}
\label{rot_lfc}
\end{center}
\end{figure}

Although the outflows are similar, the first cores look quite different between the resistive and ideal MHD models. In the ideal models, the first cores are virtually non-rotating because of the efficient magnetic braking ($R\ltsim 3 \, {\rm AU}$ in Figure~\ref{fcprof}). However, the magnetic braking is less efficient in the resistive models and there remains considerable amount of angular momentum. To show the effects of the resistivity clearly, we plot the total angular momentum within the first cores in Figure~\ref{rot_lfc}. Here we simply measure the angular momentum where the gas density is above a critical value, $\rho_{\rm crit}=10^{-13}\,{\rm g\, cm^{-3}}$. The resistivity is efficient where $\rho \gtsim 10^{-10}\, {\rm g\, cm^{-3}}$. The magnetic flux is extracted from the central high density region and redistributed to the outer thinner region, but the redistributed magnetic fields do not show any significant effect dynamically at least in this phase. Both in the models with fast and slow rotation, the first cores in the resistive models attain about twice larger angular momenta compared with the corresponding ideal MHD models. Even these differences do not have significant impact on the evolution of the first cores, they become important later in the protostellar core phase.

\begin{figure}[tb]
\begin{center}
\scalebox{1.25}{\includegraphics{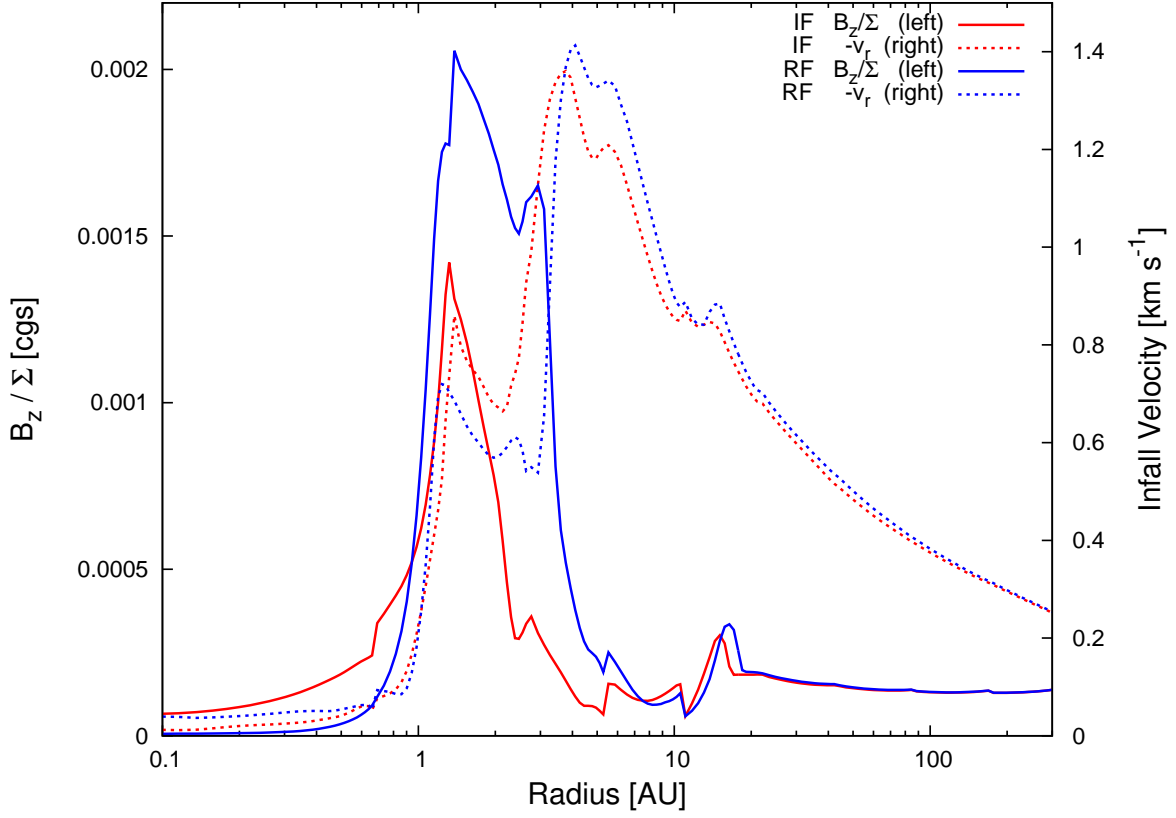}}
\caption{The ``flux-to-mass" ratio $B_z/\Sigma$ (solid) and the infall velocity $-v_r$ (dotted) of {\it IF}(red) and {\it RF} (blue) along the $x-$axis when the central density is $\rho_c \sim 4.1 \, {\rm g \, cm^{-3}}$, before the warping starts to grow. $\Sigma$ and $B_z$ is calculated in $-5\, {\rm AU} < z < 5\, {\rm AU}$ and mass-weighted average is used to evaluate $B_z$ in this region. Both models satisfy the condition for the interchange instability, $\frac{d(B_z/\Sigma)}{dr}<0$, in $1\, {\rm AU} \ltsim R \ltsim 10\, {\rm AU}$. This figure clearly shows that the magnetic fields are transported outward via Ohmic dissipation in the resistive MHD model.}
\label{fmr}
\end{center}
\end{figure}

The outstanding difference between the ideal and resistive models is that the first cores and the surrounding (pseudo) disks warp in the ideal MHD models. These are likely to be the magnetic interchange instability \citep{st90,spr95,ss01,kras12}. Uniformly-rotating disks are unstable when the ratio between the poloidal magnetic flux density and the surface density decreases radially; $\frac{d(B_z/\Sigma)}{dr}<0$. Both ideal and resistive models satisfy this condition in $1\, {\rm AU} \ltsim R \ltsim 10\, {\rm AU}$ (Figure~\ref{fmr}), though only the ideal MHD models suffer from the warping in our simulations. There are several possibilities to explain this: the timescale of growth is long in the resistive cases or we need to consider more realistic situations including the effects of finite disk thickness and gas infall to understand these phenomena, but they are beyond the scope of this paper. Possibly, the magnetically driven warping instability \citep{lai03} may also contribute to the warping. We should note that even small vertical perturbation induces artificial reconnection of magnetic fields at the mid-plane because the magnetic fields are pinched onto the disk mid-plane and the directions of the magnetic fields above and below the mid-plane are anti-parallel, then things go chaotic. Because this warping does not affect the evolution of the protostellar cores, we do not discuss it further in this work.

We can see that the gas within the dust evaporation front slowly infalls due to loss of the pressure gradient in all the models. However, the dust evaporation front is still confined in the first core and therefore its effects do not seem significant. This is different from the non-magnetized RHD simulations where the front expands beyond the first core surface \citep{sch11}. In our simulations, the angular momentum transport is very efficient and the first core properties are quite similar to the spherically symmetric cases, even in the resistive models.

Interestingly, the size (height) of the first core is slightly larger in the resistive models. It is about 3 AU in the ideal MHD models, which is similar to the spherical model, but is about 5 AU in the resistive models. This is interpreted as a consequence of energy transport and additional heating by Ohmic dissipation. The magnetic fields are transported outward, then heat up and inflate the outer region (note that the resistivity is most effective around $\rho\sim 10^{-9}\,{\rm g\,cm^{-3}}$).

\begin{figure}[tb]
\begin{center}
\scalebox{0.4}{\includegraphics{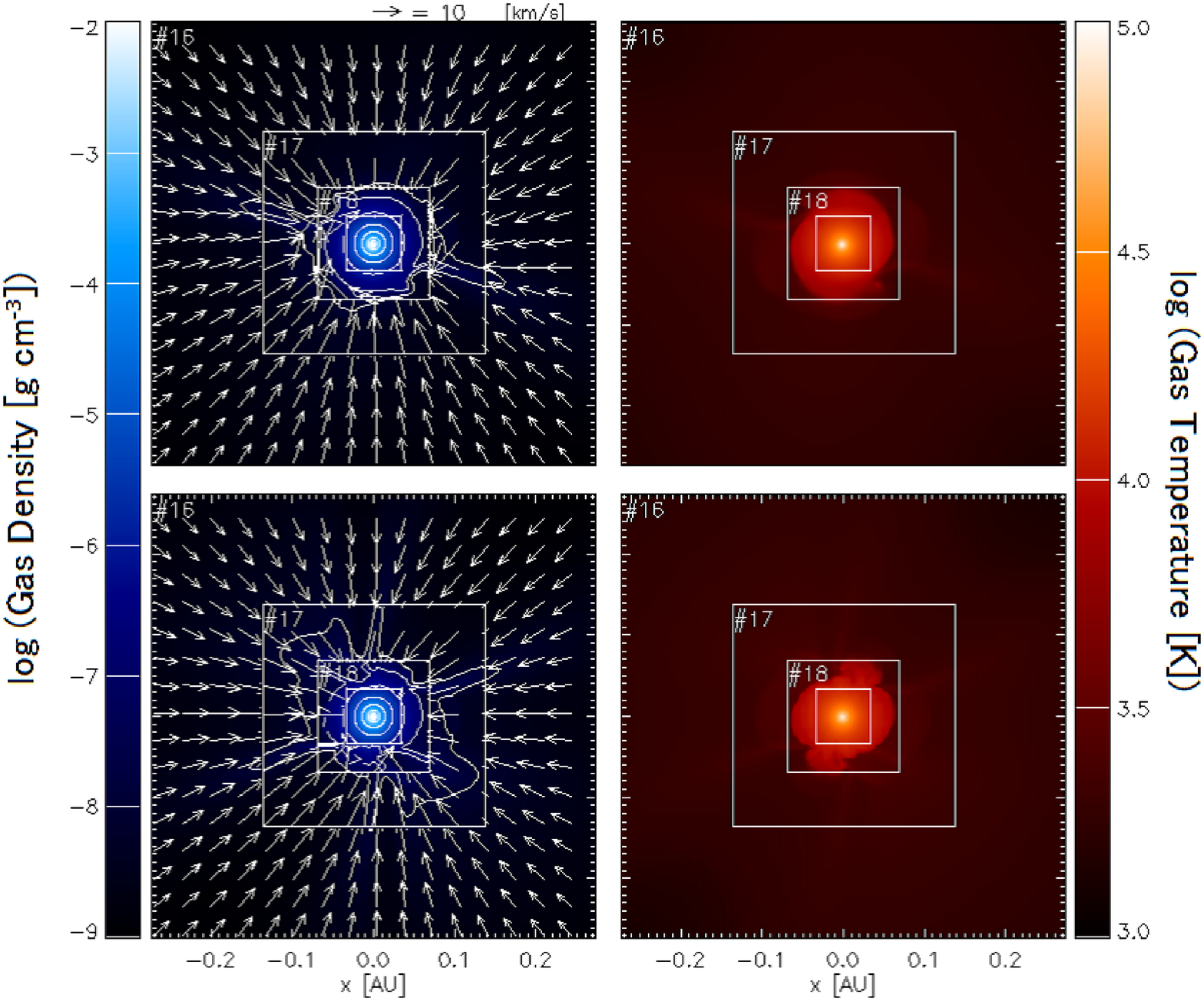}}
\caption{The vertical (top) and horizontal (bottom) cross sections of the gas density (left) and temperature (right) in the protostellar core scale ($l=16$ or $\sim 0.54\,{\rm AU}$) of Model {\it IF} at the end of the simulation.}
\label{ifsc}
\vspace{2em}
\scalebox{0.4}{\includegraphics{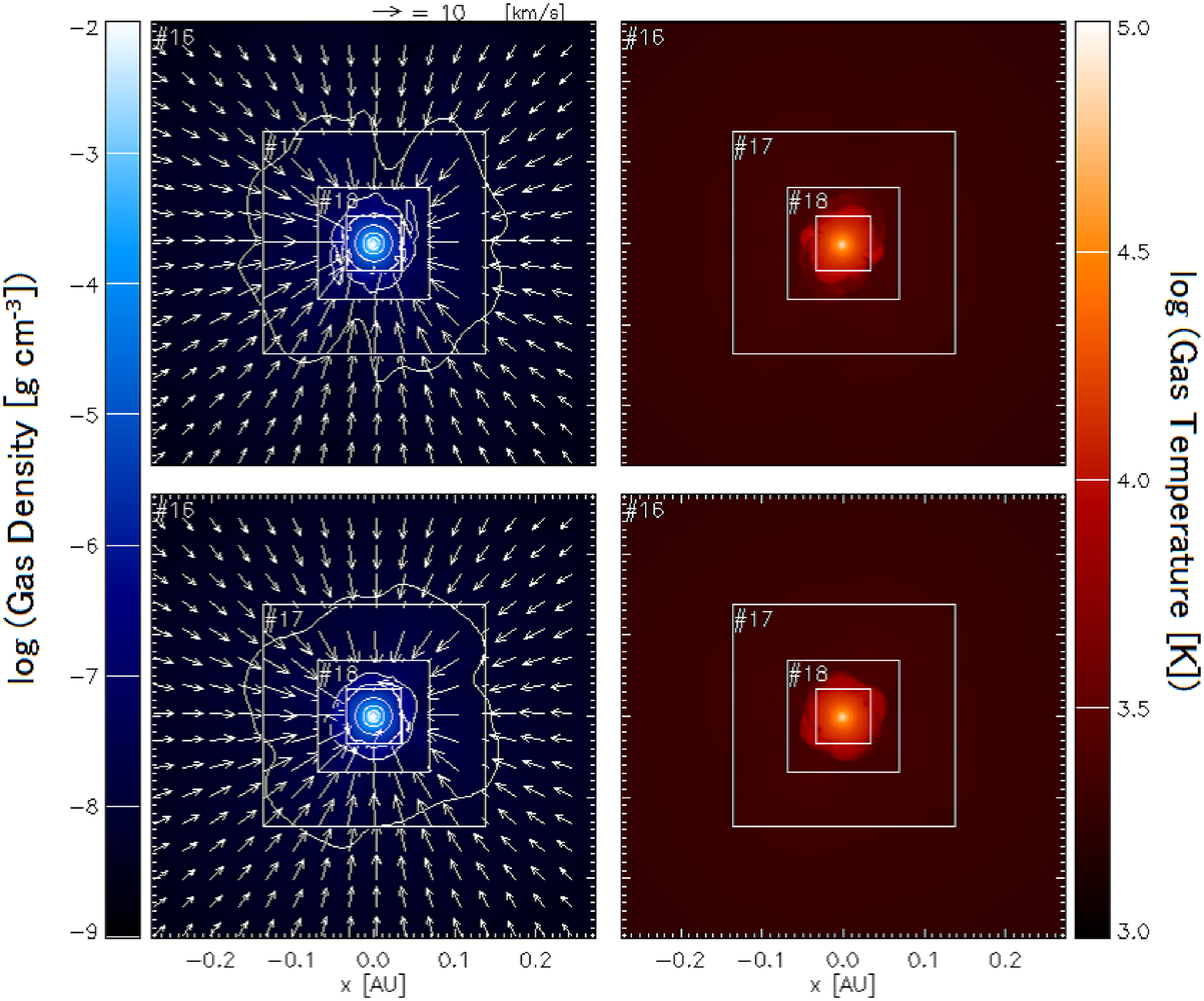}}
\caption{The same as Figure~\ref{ifsc} but of {\it IS}.}
\label{issc}
\end{center}
\end{figure}

\begin{figure}[tb]
\begin{center}
\scalebox{0.4}{\includegraphics{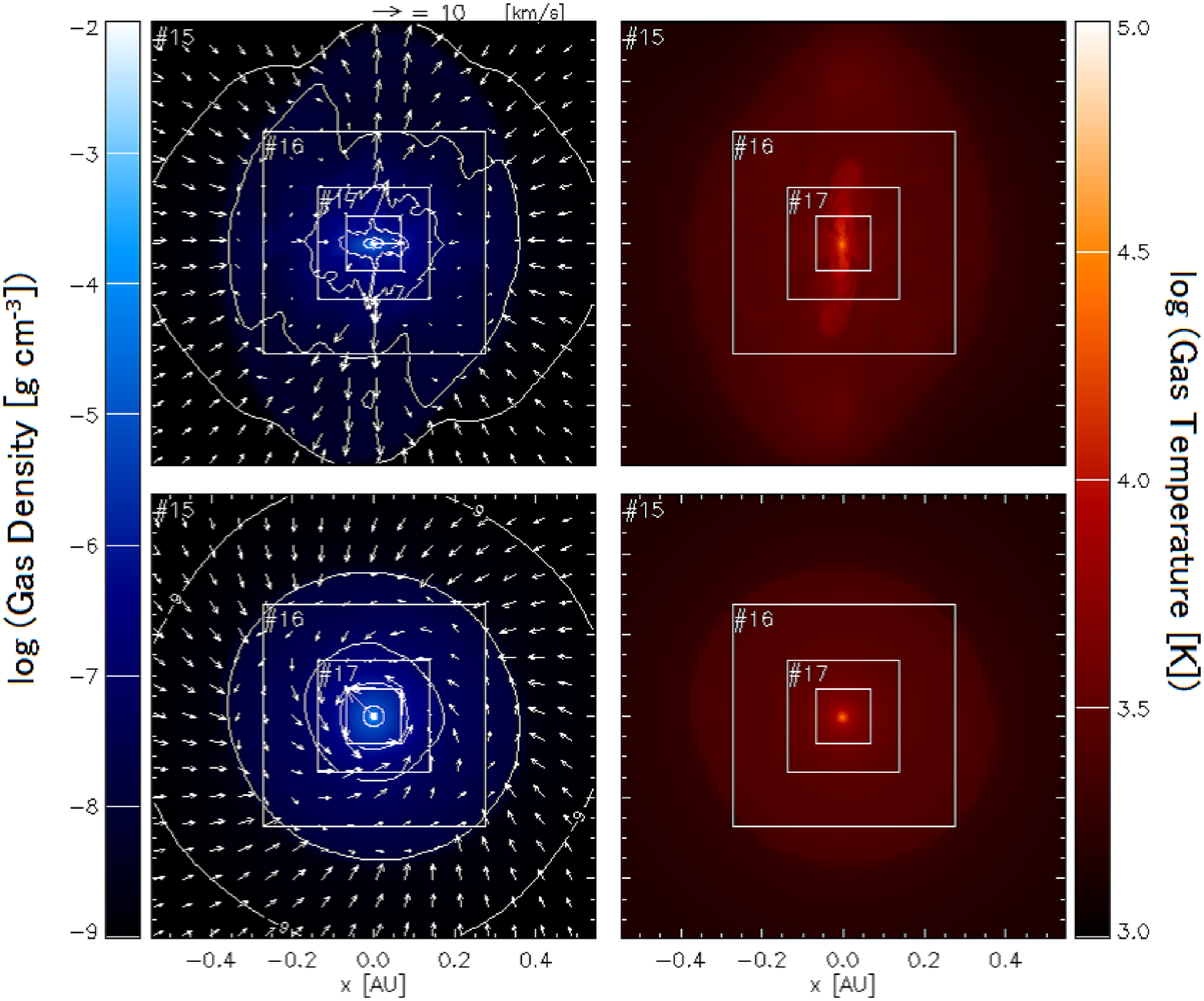}}
\caption{The same as Figure~\ref{ifsc} but of {\it RF} ($l=15$ or $\sim 1.1\,{\rm AU}$). Note that the scale is twice larger than that in Figure~\ref{ifsc}.}
\label{rfsc}
\vspace{2em}
\scalebox{0.4}{\includegraphics{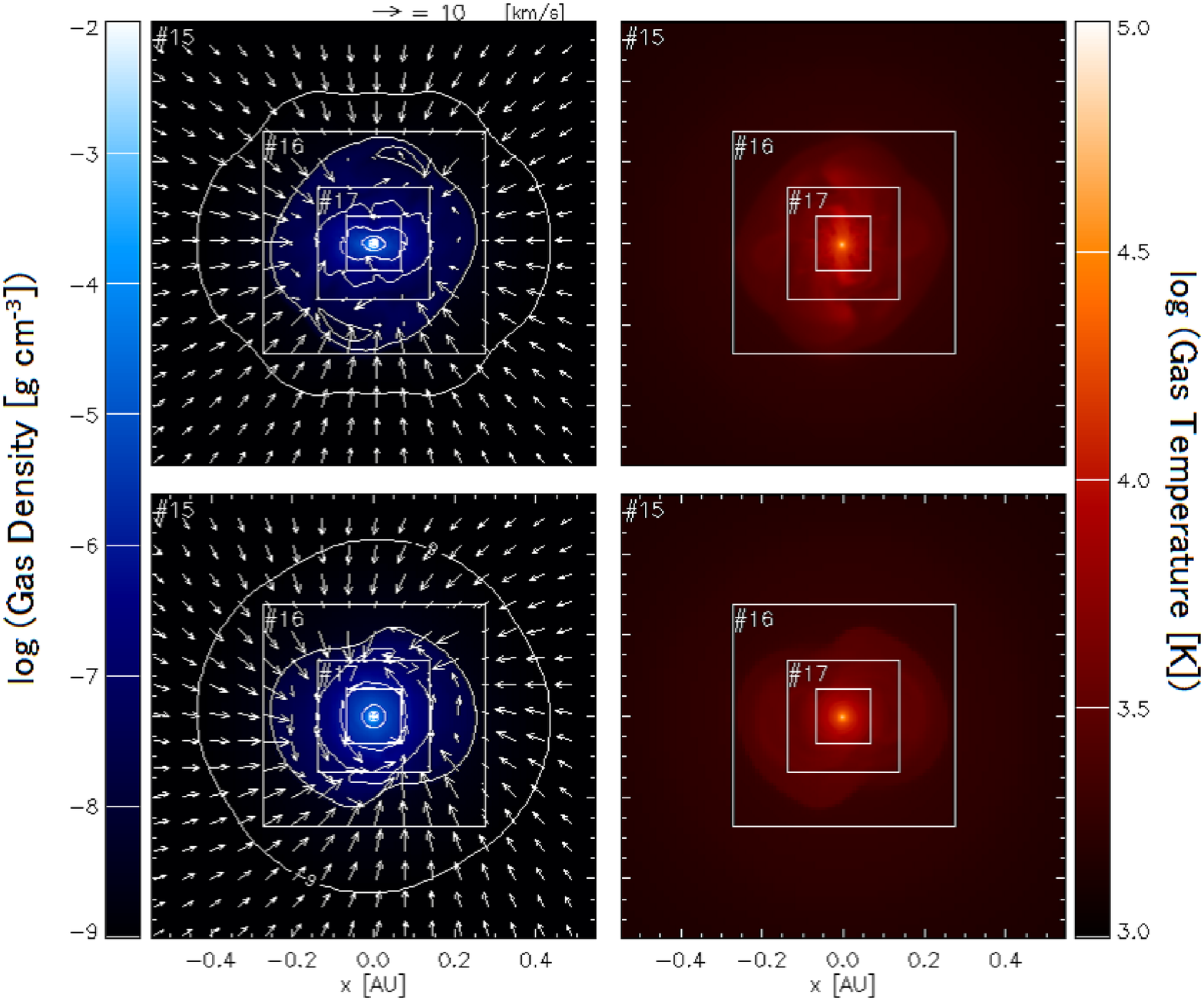}}
\caption{The same as Figure~\ref{rfsc} but of {\it RS}.}
\label{rssc}
\end{center}
\end{figure}

\clearpage
\subsubsection{Protostellar Cores and Jets}

\begin{figure}[tb]
\begin{center}
\scalebox{0.6}{\includegraphics{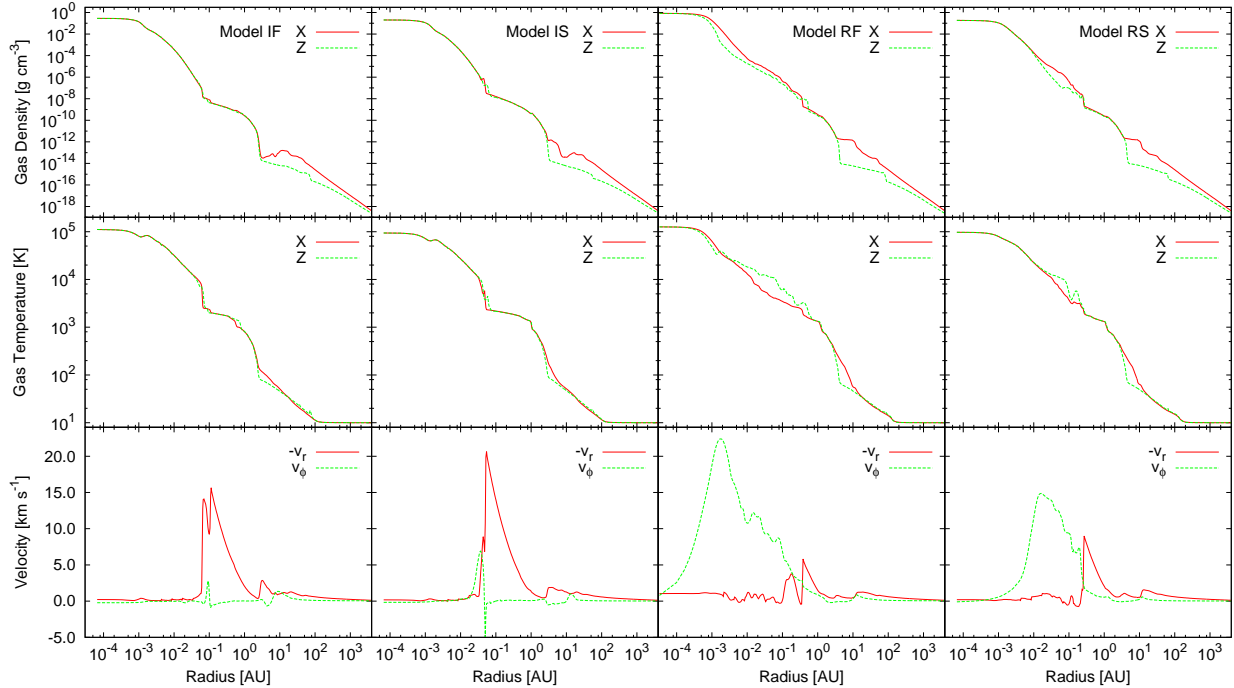}}
\caption{The same as Figure~\ref{fcprof} but at the end of the simulations.}
\label{scrprof}
\end{center}
\end{figure}

\begin{figure}[htbp]
\begin{center}
\scalebox{1.25}{\includegraphics{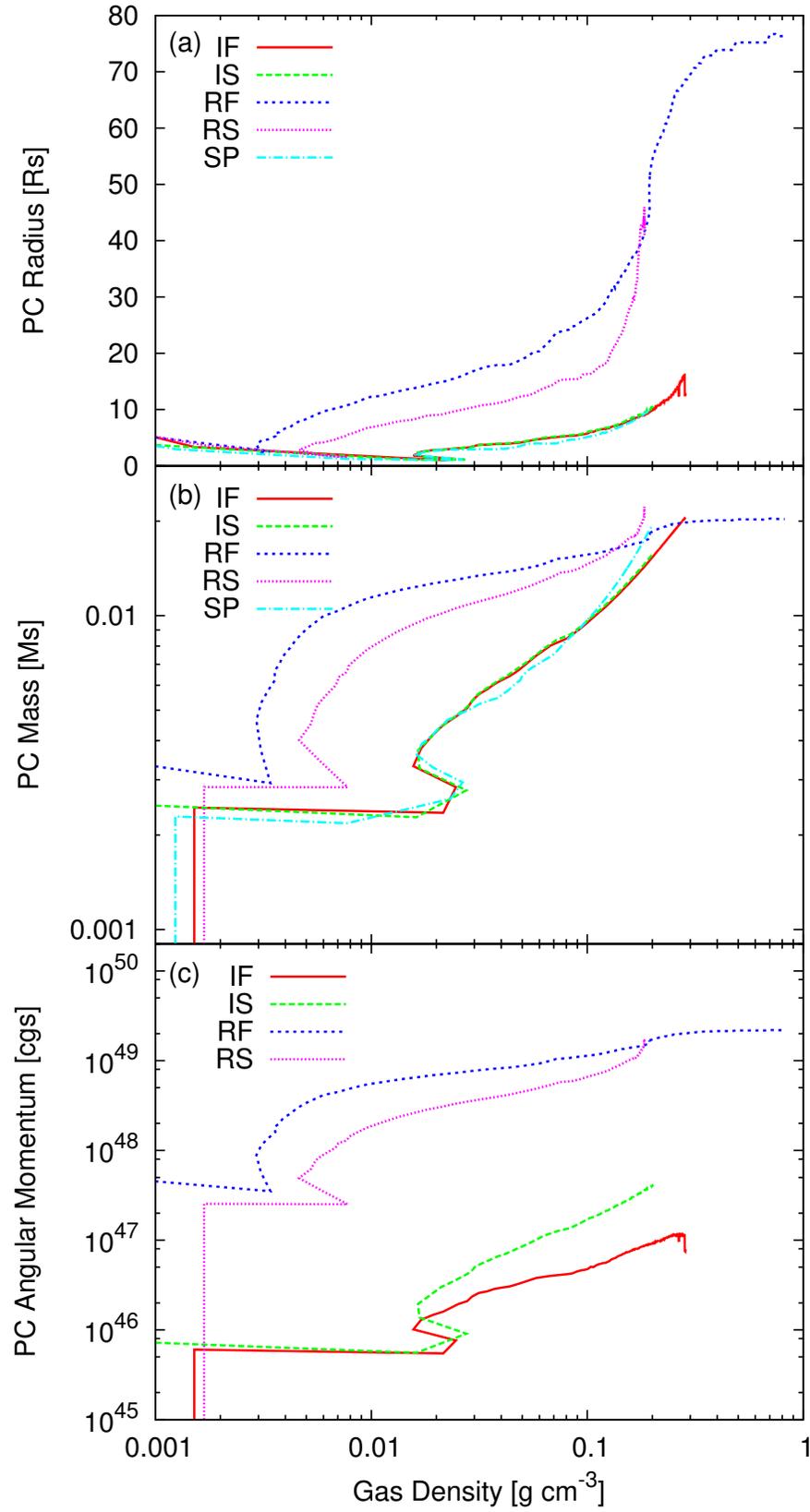}}
\caption{The evolution of the radii, masses and angular momenta of the protostellar cores as functions of the central gas density.}
\label{scprof}
\end{center}
\end{figure}

\begin{figure}[tbp]
\begin{center}
\scalebox{0.29}{\includegraphics{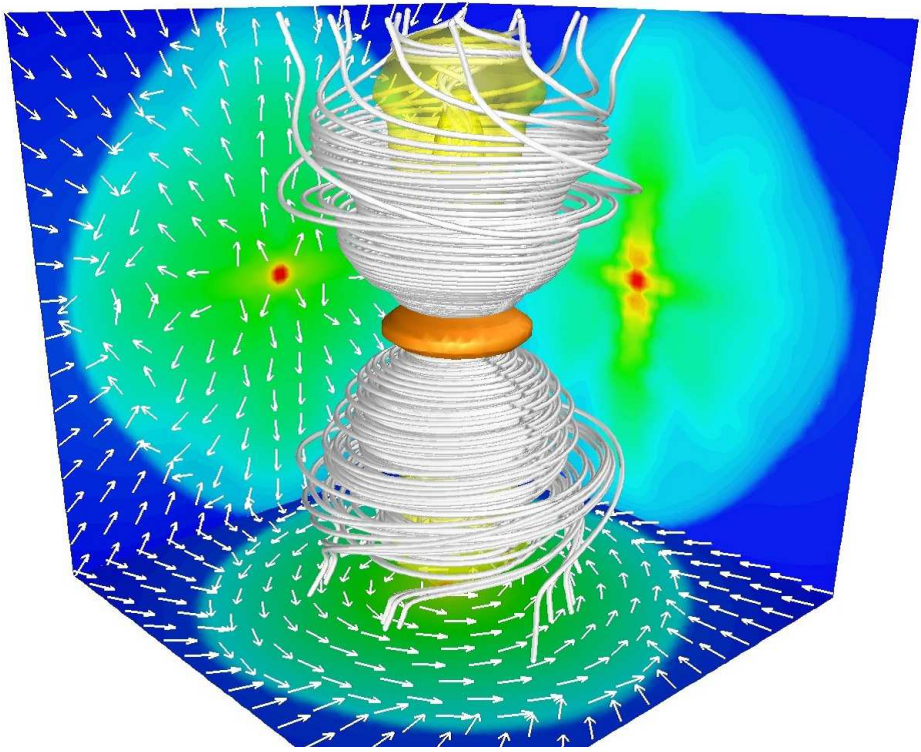}}
\caption{3D view of the protostellar core ($l=17$) in Model {\it RF}, just after the formation of the protostellar core before the growth of the kink instability. The edge of the figure is $\sim 0.27\,{\rm AU}$. The left and bottom panels are cross sections of the density and the right panel shows the temperature cross section. The high density region ($\rho > 10^{-5}\,{\rm g\, cm^{-3}}$) is visualized with the orange surface. White arrows denote the direction of the fluid motion and white lines the magnetic field lines. Fast outflowing gas ($v_z > 3\,{\rm km\, s^{-1}}$) is volume-rendered with pale yellow.}
\label{rf3d1}
\end{center}
\end{figure}

\begin{figure}[htb]
\begin{center}
\scalebox{0.29}{\includegraphics{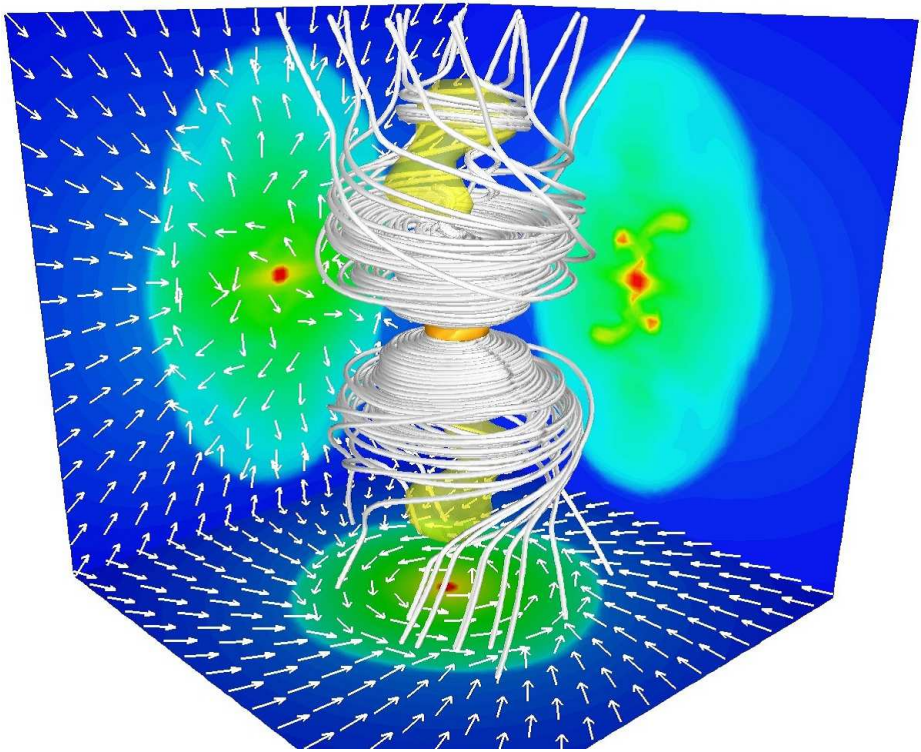}}
\caption{3D view of the protostellar core ($l=16$) in Model {\it RF}, in the growing phase of the kink instability. The edge of the figure is $\sim 0.54\,{\rm AU}$. The gas with $v_z > 4\,{\rm km\, s^{-1}}$ is rendered with pale yellow.}
\label{rf3d2}
\end{center}
\end{figure}
\begin{figure}[htb]
\begin{center}
\scalebox{0.29}{\includegraphics{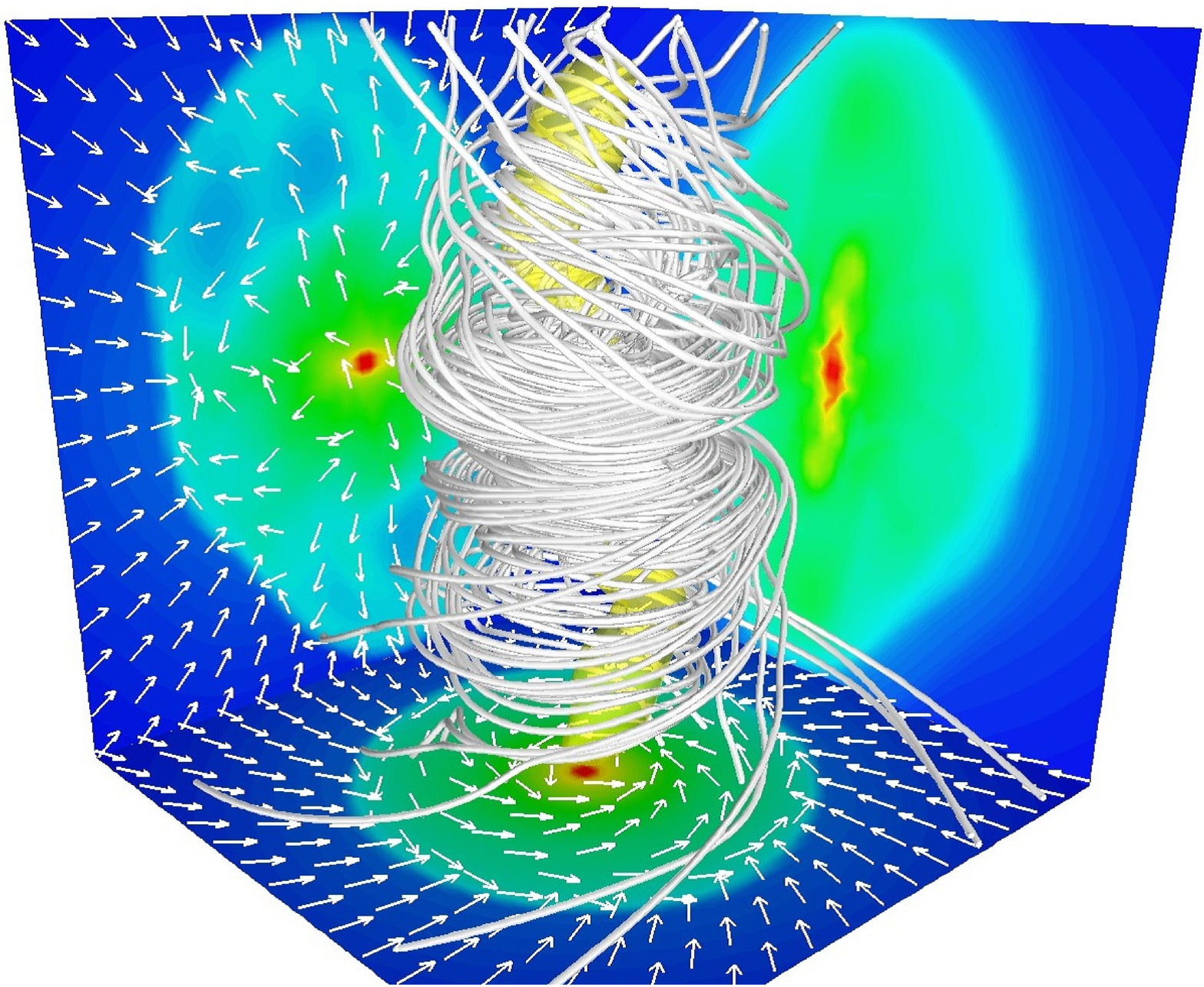}}
\caption{3D view of the protostellar core ($l=15$) in Model {\it RF} at the end of the simulation, about three months after Figure~\ref{rf3d1}. The kink instability is already grown significantly. The edge of the figure is $\sim 1.1\,{\rm AU}$. The gas with $v_z > 7\,{\rm km\, s^{-1}}$ is rendered with pale yellow.}
\label{rf3d3}
\end{center}
\end{figure}

We show the density and temperature cross sections of the the protostellar core scale ($l=15$ or $16$, corresponding to $~\sim 1.1 \, {\rm AU}$ or $\sim 0.54 \, {\rm AU}$) at the end of the simulations in Figures~\ref{ifsc} -- \ref{rssc}.  We also show the profiles along $x$- and $z$-axes in Figure~\ref{scrprof}. We stop our simulations when the central temperature reaches $T_c \sim 10^5\,{\rm K}$, corresponding to $1.05, 0.44, 0.90$ and $1.25$ years after the formation of the protostellar cores in {\it IF, IS, RF} and {\it RS}, respectively. Our simulations correspond to the earliest phases of the protostars. 

In the ideal MHD models, the properties of the protostellar cores are very similar to the spherical model. The evolution of the masses and radii of the protostellar cores are essentially identical to those in the spherical model {\it SP}. This is because the magnetic fields take almost all the angular momentum away from the gas in the central region. Contrarily, the resistive models have significantly larger angular momenta and the protostellar cores are strongly supported by rotation. To clarify the differences of the evolution between the models, we show the evolution of the radii, masses and angular momenta of the protostellar cores in Figure~\ref{scprof}. Here we define the radius of the protostellar core as the radius where the infall velocity is largest (corresponding to the shock at the surface of the protostellar core) in the $xy$-plane, and measure the mass and angular momentum within this radius. The angular momenta in the protostellar cores in the resistive models are larger than those in the ideal MHD models by more than two orders of magnitudes. The rotationally-supported protostellar cores (or ``circumstellar disks") are quickly built up within $\sim 1$ year after the formation of the protostellar cores. At the end of the simulations, the radii of the disks are about 0.35 AU in {\it RF} and 0.2 AU in {\it RS}, and they are still growing as the gas with larger angular momentum accretes. 

The protostellar cores in the resistive models also look like nearly spherical, but this is actually just a coincidence. They are supported by rotation in the horizontal direction, but they are vertically inflated due to the outflows. The toroidal magnetic fields are rapidly amplified in these rotating cores and the magnetic pressure gradient force drives the well-collimated outflows, or ``jets" (Figures~\ref{rf3d1} -- \ref{rf3d3}). The outflows are visible in the cross sections of the temperature as the hot towers. In the fast rotating model {\it RF}, the maximum outflow velocity reaches $v_z \sim 15\,{\rm km\, s^{-1}}$, while it is $\sim 6\,{\rm km\, s^{-1}}$ in {\it RS}. These velocities are comparable to the rotational velocities seen in the protostellar cores, and therefore far faster than the outflows driven from the first cores because of the deeper gravitational potential.

The magnetic fields in the rotating protostellar cores are quickly wound up and form the so-called magnetic tower. The tightly-wound magnetic fields are susceptible to the kink instability in long wavelengths. In our resistive models, the kink instability grows rapidly and the outflows start precession (Figures~\ref{rf3d1} -- \ref{rf3d3}). Although this instability disturbs the coherent toroidal magnetic fields, the outflow velocity is still getting accelerated because the bulk angular momentum in the protostellar core is increasing (Figure~\ref{scprof}) as the matter with higher angular momentum continuously accretes from the envelope, or the remnant of the first core. Therefore we expect that the outflow will be faster as the disk acquires larger angular momentum and the gravitational potential becomes deeper.

In all the models, radiation feedback from the protostellar core formation on the first core and outer envelope is not significant. The first cores seem to remain almost unaffected after the formation of the protostellar cores. This is different from previous RHD simulations without magnetic fields; the surrounding first cores are heated up by radiation and bipolar outflows are launched \citep{bate10,bate11,sch11}.  We need to study longer evolution to address this problem because we only calculate the earliest ($t\ltsim 1\,{\rm yr}$) evolution of the protostellar cores; \citet{bate10,bate11} and \cite{sch11} followed the evolution of the protostellar cores more than 15 years, longer than the free-fall time of the first cores.

\subsection{Summary of Results}

We summarize the properties of the first cores and protostellar cores at the end of the simulations in Table~\ref{results}. The table also includes the results of \citet{mi00} and \citet{mim08}. The properties of the first cores and associated outflows in the rotating models are similar to those in the previous studies. The protostellar cores are significantly larger than those in the previous studies, but they are consequences of the transient expansion which happens in the earliest evolution of the protostellar cores. This transient expansion cannot be captured with the barotropic approximation in which the shock heating is not taken into account. The protostellar cores acquire their masses very quickly in this phase ($\sim 0.02 \,M_\odot$ in a year). Because this phenomenon has very short time scale (estimated to be $\sim 5$ years), our models are not inconsistent with \citet{mi00}, and also consistent with the present-day case of \citet{ohy10}. This expansion will affect the properties of the protostellar core when the core settles down after this expansion. It may be critically important in the further evolution of the protostars as the initial condition. However, to confidently discuss this phenomenon and the properties of the protostellar cores, we have to calculate the evolution far longer.

The protostellar cores formed in the ideal MHD models are essentially non-rotating because of the efficient angular momentum transport via magnetic fields. In the resistive models, on the other hand, the protostellar cores attain considerably large angular momenta and the rotationally-supported disks emerge in their earliest phases, and they will continuously evolve into circumstellar disks. Although it is difficult to distinguish the disk from the central protostar in our simulations at this phase, they are going to separate into a thermally-supported protostar and a rotationally-supported disk when radiation cooling takes place and reduces the thermal support. Thus Ohmic dissipation remedies the magnetic braking catastrophe.

Despite the significant difference of the thermal evolution in the protostellar cores, the properties of the outflows and circumstellar disks associated with the protostellar cores in the resistive models are consistent with the previous MHD studies using the barotropic approximation \citep{mim08}. This is because they are mainly determined by the interaction between magnetic fields and rotation, and the effects of the thermal evolution have less significant impacts on their properties.

\begin{table}[tbp] 
\begin{center}
\begin{tabular}{c|cccccc}
Model & $\tau_{\rm FC}({\rm yrs})$ & $R_{\rm FC}({\rm AU})$ & $D_{\rm OF}({\rm AU})$ & $R_{\rm PC}({\rm R_\odot})$ & $M_{\rm PC}({\rm M_\odot})$ & $V_{\rm Jet}\,({\rm km\, s^{-1}})$\\
\hline
{\it SP} & 650 & 3 & 0 & $10^*$ & 0.02 & 0\\
{\it IS} & 720 & 3 & 55 & $10^*$ & 0.02 & 0 \\
{\it IF} & 800 & 3 & 70 & $17^*$ & 0.02 & 0 \\
{\it RS} & 850 & 5 & 60 & 45 & 0.02 & 5 \\
{\it RF} & 950 & 5 & 80 & 75 & 0.02 & 15 \\
\hline
{\it MI} & 650 & 3 & 0 & $4$ & 0.016 & 0\\
{\it MR}$^\dagger$ & -- & 0.5 & -- & $8.2$ & 0.008 & 15
\end{tabular}
\end{center}
\caption{Summary of the properties of the first cores and the protostellar cores. From left to right: the lifetime of the first core, the thermally-supported radius of the first core, the distance the outflow traveled during the first core lifetime, the thermally- or rotationally-supported radius of the protostellar core, the mass of the protostellar core and the maximum velocity of the jet driven from the protostellar core. The quantities marked with $^*$ are strongly affected by the transient expansion. For comparison, we also show the results of \citet{mi00} ({\it MI}), the snapshot labeled ``10" which corresponds to 19 years after the second collapse, and the model {\it MR} of \citet{mim08}. $\dagger$: We adopt their disk radius as the protostellar core radius and we show the total mass of the protostar, disk and outflows because we do not distinguish them in this study. The size of the thermally supported protostellar core in {\it MR} is small, $\sim 1 R_\odot$, because of the barotropic approximation.}
\label{results}
\end{table}

\section{Conclusions and Discussions}

We performed 3D nested-grid RMHD simulations of formation of protostellar cores from molecular cloud cores with and without Ohmic dissipation, and revealed the earliest evolution (only 1 year after the formation) of the protostellar cores. These simulations are, to our knowledge, the first 3D RMHD simulations in the world following the whole evolution from molecular cloud cores to protostellar cores with realistic physics. We successfully revealed the realistic evolution in the early phase of protostellar collapse.

Preceding studies \citep{tomida10a,tomida10b,bate11} had showed that the barotropic approximation fails to reproduce the realistic thermal properties, and here we also revealed that the discrepancy is more prominent in the protostellar core phase, because the barotropic approximation does not take account of the shock heating and therefore tend to underestimate the temperature, which results in the smaller radii of the first and protostellar cores and cannot reproduce the transient expansion of the protostellar cores. In order to capture the realistic thermal evolution, RMHD simulations are thus essential. Our results can be used as the initial and boundary conditions for the stellar evolution simulations \citep[e.g.,][]{bcg,hok11}.

We found two qualitatively different phenomena between the ideal MHD models and the resistive MHD models: formation of circumstellar disks and outflows from protostellar cores. In the resistive models, the circumstellar disks are quickly built up in the vicinity of the formed protostars. Although the formed disks are still very small ($R < 0.35 \, {\rm AU}$), we expect the disks will grow continuously as the gas with larger angular momentum accretes. \citet{mim11b} performed resistive MHD simulations with the sink cell technique and the barotropic approximation and showed that the rotationally-supported disk grows to $R > 200 \, {\rm AU}$ while the high density region where Ohmic dissipation works expands as the disk grows. The two-component outflows are launched from different scales as the natural by-products: the slow loosely-collimated outflow driven from the first core and the fast well-collimated jet driven from the protostellar core. These are consistent with previous studies using the barotropic approximation \citep{mim06,mim07,mim08}. Although the maximum velocity of the outflow is still not very fast ($v_z\sim 15\,{\rm km\, s^{-1}}$), we expect that it will get faster as the protostellar core grows. Meanwhile, we expect that the outflow from the first core is continuously driven from the outer disk, or the remnant of the first core until the accretion stops \citep{mm11,mim11b}. If it is the case, our simulations can naturally explain the observed high-velocity jets, especially the multi-component outflows \citep[for example,][]{cfl00,sg09}. In the ideal MHD cases, on the other hand, the protostellar cores are almost the same with the spherically symmetric case and no outflow is launched as a result of efficient magnetic braking. These results do not necessarily mean that circumstellar disks are never formed in ideal MHD simulations. The circumstellar disks and outflows may be formed later when the gas with large angular momentum accretes sufficiently, because the magnetic braking is not a process which reduces the total angular momentum but transports the angular momentum from the disk to the thin envelope within the molecular cloud via Alfv\'en waves. To simulate such a system, long-term simulations with reasonable boundary conditions are required.

Although ``when and how circumstellar disks are formed?" is a crucial question, there has been no conclusive answer. Our results suggest that a circumstellar disk is formed in the earliest phase of protostar evolution, essentially in parallel. \citet{jor09} suggested that Class-0 sources (young, embedded) are associated with more massive disks than Class-I sources, which support the early formation of circumstellar disks. We need more elaborate observations with higher spatial resolution to answer this question, and we expect the Atacama Large Millimeter/submillimeter Array (ALMA) will provide good clues to this problem.

In our simulations, the formation of the protostellar core does not affect the first core and outer envelope. Considering the large energy released in the second collapse and subsequent accretion with the high accretion rate, we expect that the feedback like \citet{bate11} and \citet{sch11} reported may happen; bipolar outflows may be launched by radiation heating from the protostellar core. There are two possible interpretations of the discrepancy: magnetic fields and duration of the simulations. Because of the efficient angular momentum transport due to magnetic fields, the first cores are not supported by rotation even in the resistive models and the structures of the accretion flows are completely different between the models with and without magnetic fields. Indeed, \citet{bate11} showed that the outflows are less energetic in the models with slower rotation. Therefore magnetic fields possibly can prevent formation of such outflows driven by radiation heating from the protostellar cores.

The other point is that we only calculate about 1 year after the protostellar core formation while \citet{bate11} and \citet{sch11} followed the evolution longer than 15 years. Because 1 year is shorter than the free-fall time of the first core, the protostellar cores are still deeply obscured in the remnants of the first cores in our simulations. If this is the case, how long will it take for the feedback to become prominent? As we mentioned above, the answer is the free-fall timescale of the first core, which is about 5 years. Another interesting factor is the interaction between the first core and the outflow from the protostellar core. When the outflow from the protostellar core penetrates the first core, it will have some influence on the first core.  Assuming the constant traveling velocity of $15 \,{\rm km\,s^{-1}}$, it will reach the surface of the first core in about 1.5 years. Unfortunately, it is almost hopeless to follow such a long evolution with our current code because the simulation timesteps are too small, less than 1 minute, and are still getting smaller. In order to perform the long-term simulations, we have to modify our code, for example, introducing the sink particle technique \citep{bate95,krm04,fed10} to replace the protostellar core with a subgrid stellar evolution model \citep[e.g.,][]{bc02,yb08,ho09}. To construct reasonable subgrid models, our results can be used as templates. Since the radiation feedback from the formed protostar can be highly anisotropic \citep[``flashlight effect",][]{yb99} due to the small-scale optically-thick disk \citep{vaidya,tn11}, high resolution down to the protostellar core scale is of crucial importance even when we introduce the sink particle and subgrid stellar evolution model. 

All the physical processes involved in this work play crucial roles in star formation processes. Magnetic fields efficiently extract the angular momentum and drive the outflows from the first cores. Ohmic dissipation suppresses the angular momentum transport which is too efficient in the ideal MHD models and enables the formation of the circumstellar disks and the fast outflows from the protostellar cores. The EOS and radiation transfer give the realistic thermal evolution and enable us to discuss the dynamics and thermodynamics quantitatively, which cannot be achieved with the barotropic approximation. We should note, however, that there is no established test for such a complicated system, although we tested each part of our code separately with standard tests. To verify the validity of the codes, we are planning to perform comparisons with other groups. \\

We thank Prof.\ Matthew R. Bate, Prof.\ Shu-ichiro Inutsuka, Prof.\ Kazuyuki Omukai and Dr.\ Takashi Hosokawa for fruitful discussions. We also thank Prof. James M. Stone for reading the manuscript carefully. We are grateful to Prof.\ Masahiro Ikoma for providing his notes on EOS and Dr.\ Takahiro Miyoshi for providing information about the HLLD and HLLD$-$ solvers. We also appreciate Mr.~Richard B. Reichart checking the manuscript and improving English. Numerical computations were partly performed on NEC SX-9 at Center for Computational Astrophysics of National Astronomical Observatory of Japan, at Japan Aerospace Exploration Agency and at Osaka University. This work is partly supported by the Ministry of Education, Culture, Sports, Science and Technology (MEXT), Grants-in-Aid for Scientific Research, 21244021 (Tomisaka) and 23540270 (Matsumoto). K.~Tomida, Y.~Hori and S.~Okuzumi are supported by the Research Fellowship from the Japan Society for the Promotion of Science (JSPS).

\section*{Appendix 1. Equation of State}
In some previous studies, the simple EOS of perfect gas has been often used, i.e., the adiabatic index $\Gamma$ is assumed to be a constant throughout simulations, $7/5$ which is valid for completely idealized diatomic molecular gas, or $5/3$ which is correct only when the gas temperature is very low. However, the adiabatic index is not constant in reality. In the low temperature molecular cloud ($T\ltsim 100 \,{\rm K}$), molecular hydrogen behaves like monoatomic gas, $\Gamma\sim 5/3$, because the collisional energy is insufficient to excite the rotational degrees of freedom and only translations are excited. When the gas temperature exceeds $T\gtsim 100\, {\rm K}$, the rotational degrees start to be excited and contribute to the heat capacity. Then the adiabatic index decreases, $\Gamma\sim 7/5$, close to that of the diatomic gas. The adiabatic index is of critical importance in the thermal evolution of the gas, and also in the stability of the gas against the gravitational instability \citep{stm09}; the stiffer gas (with larger $\Gamma$) is more stable gravitationally because it reacts more strongly against compression. Therefore we need to consider these quantum thermodynamic behaviors in our EOS.

The second collapse is driven by the endothermic reaction of hydrogen molecule dissociation. In order to simulate the evolution in the second collapse phase, we need to take the chemical reactions into account. However, it requires quite a large computational cost to solve the network of chemical reactions in every fluid element while solving (radiation) hydrodynamics. Fortunately, because we are mainly interested in dense gas, we can assume that the timescale of chemical reactions is shorter than the dynamical timescale. Therefore we implement the chemical reactions related to major species within the EOS on the assumption of the local thermodynamic and chemical equilibrium.

We require some assumptions to calculate the EOS for simplicity:
\begin{itemize}
\setlength{\parskip}{0cm}
\setlength{\itemsep}{0cm}
\item The gas is in the local thermodynamic and chemical equilibrium (except for the ortho/para ratio of molecular hydrogen; see below).
\item All the atoms, molecules and ions are in the ground states.
\item The EOS can be calculated by simple summation of each component and each degree of freedom, i.e., the interactions between components and other non-ideal effects are neglected.
\item Only seven major species (${\rm H_2, H, H^+, He, He^+, He^{2+}}$ and ${\rm e^-}$) are considered and other elements are neglected.
\end{itemize}
Based on these assumptions, we calculate the EOS using the statistical mechanics theory. Here we also assume that the gas has the solar abundance, $X=0.7$ and $Y=0.28$.

\subsection*{Partition Functions}
Here we describe partition functions of each element. The partition function of a species $i$ can be divided into five parts; translation $Z_{{\rm tr},i}$, rotation $Z_{{\rm rot},i}$, vibration $Z_{{\rm vib},i}$, spin $Z_{{\rm spin},i}$ and electron excitation $Z_{{\rm elec},i}$ (we include contributions from ${\rm H_2}$ dissociation and ionization of hydrogen and helium in the electron excitation part):
\begin{eqnarray}
Z_i=V\times Z_{{\rm tr},i}\times Z_{{\rm rot},i}\times Z_{{\rm vib},i} \times Z_{{\rm spin},i} \times Z_{{\rm elec},i},
\end{eqnarray}
where V is the volume. We calculate these partition functions by the standard procedure. In the following descriptions, $m_i$ denotes the mass of $i$-species, $h$ the Planck constant and $k$ the Boltzmann constant, respectively. The partition functions not explicitly described are unity.\\

\noindent Molecular hydrogen:
\begin{eqnarray}
Z_{\rm tr,H_2}&=&\frac{(2\pi m_{\rm H_2}kT)^{3/2}}{h^3},\\
Z_{\rm rot,H_2}&=&\left(Z_{\rm rot,H_2}^{\rm even}\right)^{\frac{1}{4}}\left[3 Z_{\rm rot,H_2}^{\rm odd}\exp\left(\frac{\theta_{\rm rot}}{T}\right)\right]^{\frac{3}{4}},\\
Z_{\rm rot,H_2}^{\rm even}&=&\sum_{j=0,2,4,...}(2j+1)\exp\left[-\frac{j(j+1)\theta_{\rm rot}}{2T}\right],\\
Z_{\rm rot,H_2}^{\rm odd}&=&\sum_{j=1,3,5,...}(2j+1)\exp\left[-\frac{j(j+1)\theta_{\rm rot}}{2T}\right],\\
Z_{\rm vib,H_2}&=&\frac{1}{2\sinh{(\theta_{\rm vib}/2T)}},\\
Z_{\rm spin,H_2}&=&\left(2\cdot \frac{1}{2}+1\right)^2=4,\\
Z_{\rm elec,H_2}&=&2,
\end{eqnarray}
where $\theta_{\rm rot}=170.64\, {\rm K}$ is the excitation temperature of rotation and $\theta_{\rm vib}=5984.48\, {\rm K}$ is that of vibration. Molecular hydrogen has two forms: orthohydrogen with aligned nuclear spins and odd rotational states, and parahydrogen with antiparallel nuclear spins and even rotational states. Here we assumed the ratio of orthohydrogen to parahydrogen is 3:1. This ratio has significant impact on the dynamics of collapsing molecular cloud cores in the relatively low temperature region because thermodynamic properties related to rotation of molecular hydrogen depend on the nuclear spins and rotational states. Unfortunately this ratio in star forming regions is quite unclear because of observational difficulties. Although parahydrogen is more stable, some observations of interstellar dark clouds suggest that the ratio is considerably far from the equilibrium value even in the cold environment; \citet{pagani11} proposed that the ortho/para ratio is larger than 0.1. On the other hand, \citet{dislaire} claimed that the ratio is quite small, $\sim 10^{-3}$. In this work, we calculate the EOS using the ortho/para ratio of 3:1. This assumption helps us interpret our simulation results because the adiabatic index $\Gamma$ decreases monotonically \citep{boley07} and also compare our results with recent simulations performed by \citet{bate10,bate11} (but \citet{stm09} assumed the equilibrium ratio).\\

\noindent Atomic hydrogen:
\begin{eqnarray}
Z_{\rm tr,H}&=&\frac{(2\pi m_{\rm H}kT)^{3/2}}{h^3},\\
Z_{\rm spin,H}&=&2\cdot\frac{1}{2}+1=2,\\
Z_{\rm elec,H}&=&2\exp\left(-\frac{\chi_{\rm dis}}{2kT}\right),
\end{eqnarray}
where $\chi_{\rm dis}=7.17\times 10^{-12} \, {\rm erg}$ is the dissociation energy of ${\rm H_2}$ \citep{liu09}.\\

\noindent Ionized hydrogen:
\begin{eqnarray}
Z_{\rm tr,H^+}&=&\frac{(2\pi m_{\rm H^+}kT)^{3/2}}{h^3},\\
Z_{\rm spin,H^+}&=&2\cdot\frac{1}{2}+1=2,\\
Z_{\rm elec,H^+}&=&2\exp\left(-\frac{\chi_{\rm dis}+2\chi_{\rm ion}}{2kT}\right),
\end{eqnarray}
where $\chi_{\rm ion}=2.18\times 10^{-11} \, {\rm erg}$ is the ionization energy of atomic hydrogen.\\

\noindent Atomic and ionized helium:
\vspace{-1em}
\begin{eqnarray}
Z_{\rm tr,He}&=&\frac{(2\pi m_{\rm He}kT)^{3/2}}{h^3},\\
Z_{\rm tr,He^+}&=&\frac{(2\pi m_{\rm He^+}kT)^{3/2}}{h^3},\\
Z_{\rm elec,He^+}&=&\exp\left(-\frac{\chi_{\rm He,1}}{kT}\right),\\
Z_{\rm tr,He^{2+}}&=&\frac{(2\pi m_{\rm He^{2+}}kT)^{3/2}}{h^3},\\
Z_{\rm elec,He^{2+}}&=&\exp\left(-\frac{\chi_{\rm He,1}+\chi_{\rm He,2}}{kT}\right).
\end{eqnarray}
where $\chi_{\rm He,1}=3.94\times 10^{-11} \, {\rm erg}$ and $\chi_{\rm He,2}=8.72\times 10^{-11} \, {\rm erg}$ are the first and second ionization energies of helium.\\

\noindent Electron:
\begin{eqnarray}
Z_{\rm tr,e}&=&\frac{(2\pi m_{\rm e}kT)^{3/2}}{h^3},\\
Z_{\rm spin,e}&=&2\cdot\frac{1}{2}+1=2.
\end{eqnarray}

\subsection*{Chemical Reactions and Number Densities}
The grand canonical partition function of each species is defined as:
\begin{eqnarray}
\Theta_i(\mu_i,V,T)=\sum_{N_i}\exp\left(\frac{N_i\mu_i}{kT}\right)\frac{Z_i^{N_i}}{N_i!}=\exp\left[Z_i\exp\left(\frac{\mu_i}{kT}\right)\right],
\end{eqnarray}
where $\mu_i$ is the chemical potential of $i$-species and $N_i$ is the total number of $i$-species. The grand potential can be derived from the grand canonical partition function:
\begin{eqnarray}
\Omega_i(\mu_i,V,T)=-kT\log\Theta_i=-kTZ_i\exp\left(\frac{\mu_i}{kT}\right).
\end{eqnarray}
The total grand potential can be calculated from the summation of each component:
\begin{eqnarray}
\Omega=\sum_i\Omega_i.
\end{eqnarray}

We calculate required thermodynamic variables from these functions. First, we calculate the number density of each species based on the chemical equilibrium. The number density of $i$-species is derived from the partial derivative of the grand potential with respect to $\mu_i$:
\begin{eqnarray}
n_i=\frac{1}{V}\left(\frac{\partial \Omega}{\partial \mu_i}\right)_{\mu_j,V,T}=z_i\exp\left(\frac{\mu_i}{kT}\right),
\end{eqnarray}
where $z_i=Z_i/V$. This relation yields:
\begin{eqnarray}
\mu_i=kT\log\frac{n_i}{z_i}.
\end{eqnarray}

We consider only four reactions between the seven species dominant in relatively dense (but not too dense) gas for star formation problems:
\begin{eqnarray}
{\rm H_2} & \longleftrightarrow &  {\rm 2H},\\
{\rm H} & \longleftrightarrow &  {\rm H^++e^-},\\
{\rm He} & \longleftrightarrow &  {\rm He^++e^-},\\
{\rm He^+} & \longleftrightarrow &  {\rm He^{2+}+e^-}.
\end{eqnarray}
Then the number densities can be calculated from the balance between the chemical potentials in these chemical reactions. 
\begin{eqnarray}
\mu_{\rm H_2}=2\mu_{\rm H} & \Longrightarrow &  \frac{n_{\rm H}^2}{n_{\rm H_2}}=\frac{z_{\rm H}^2}{z_{\rm H_2}}\equiv K_{\rm dis},\label{chm1}\\
\mu_{\rm H}=\mu_{\rm H^+}+\mu_e & \Longrightarrow &  \frac{n_{\rm H^+}n_e}{n_{\rm H}}=\frac{z_{\rm H^+}z_e}{z_{\rm H}}\equiv K_{\rm ion},\\
\mu_{\rm He}=\mu_{\rm He^+}+\mu_e & \Longrightarrow &  \frac{n_{\rm He^+}n_e}{n_{\rm He}}=\frac{z_{\rm He^+}z_e}{z_{\rm He}}\equiv K_{\rm He,1},\\
\mu_{\rm He^+}=\mu_{\rm He^{2+}}+\mu_e & \Longrightarrow &  \frac{n_{\rm He^{2+}}n_e}{n_{\rm He^+}}=\frac{z_{\rm He^{2+}}z_e}{z_{\rm He^+}}\equiv K_{\rm He,2}.\label{chm4}
\end{eqnarray}
The RHS term of each equation, $K_{*}$, can be calculated from the partition functions. We have three additional relations; conservation of the total number density of hydrogen, conservation of the total number density of helium and the charge neutrality:
\begin{eqnarray}
2n_{\rm H_2}+n_{\rm H}+n_{\rm H^+}&=&n_{\rm tot}^{\rm H}\hspace{0.5em}\left(=\frac{\rho X}{m_{\rm H}}\right),\label{chm5}\\
n_{\rm He}+n_{\rm He^+}+n_{\rm He^{2+}}&=&n_{\rm tot}^{\rm He}\hspace{0.5em}\left(=\frac{\rho Y}{m_{\rm He}}\right),\label{chm6}\\
n_{\rm H^+}+n_{\rm He^+}+2n_{\rm He^{2+}}&=&n_e.\label{chm7}
\end{eqnarray}
We eliminate $n_{\rm H_2}, n_{\rm H^+}, n_{\rm He^+}$ and $n_{\rm He^{2+}}$ from (\ref{chm5} -- \ref{chm7}) using (\ref{chm1} -- \ref{chm4}):
\begin{eqnarray}
2\frac{n_{\rm H}^2}{K_{\rm dis}}+n_{\rm H}+\frac{n_{\rm H}}{n_e}K_{\rm ion}&=&n_{\rm tot}^{\rm H},\label{chm8}\\
n_{\rm He}\left(1+\frac{K_{\rm He,1}}{n_e}+\frac{K_{\rm He,1}K_{\rm He,2}}{n_e^2}\right)&=&n_{\rm tot}^{\rm He},\label{chm9}\\
\frac{n_{\rm H}}{n_e}K_{\rm ion}+\frac{n_{\rm He}}{n_e}K_{\rm He,1}+2\frac{n_{\rm He}}{n_e^2}K_{\rm He,1}K_{\rm He,2}&=&n_e.\label{chm10}
\end{eqnarray}
By substituting $n_{\rm H}$ and $n_{\rm He}$ (we can determine the solution of (\ref{chm8}) uniquely since all the physical variables are positive) to (\ref{chm10}), we obtain one non-linear equation of $n_e$:
\begin{eqnarray}
f(n_e)=\frac{2n_e^2n_{\rm tot}^{\rm H}K_{\rm ion}}{\sqrt{(n_e+K_{\rm ion})^2+\frac{8}{K_{\rm dis}}n_{\rm tot}^{\rm H}n_e^2}+(n_e+K_{\rm ion})}\nonumber\\
+\frac{K_{\rm He,1}n_e+2K_{\rm He,1}K_{\rm He,2}}{n_e^2+K_{\rm He,1}n_e+K_{\rm He,1}K_{\rm He,2}}n_{\rm tot}^{\rm He}n_e^2-n_e^3=0.
\end{eqnarray}
Then we solve this equation numerically using the bi-section method 
 (note that the first term of $f(n_e)$ is already modified to avoid the round-off error). Using the obtained $n_e$, it is straightforward to calculate the number density of each species.

In order to derive thermodynamic variables, the temperature and density derivatives of the number densities are required. We take logarithmic differentiation of (\ref{chm1} -- \ref{chm7}), then they yield:
\begin{eqnarray}
\left(\hspace{-1ex}
\begin{array}{ccccccc}
1&-2&0&0&0&0&0\\
2n_{\rm H_2}&n_{\rm H}&n_{\rm H^+}&0&0&0&0\\
0&-1&1&1&0&0&0\\
0&0&n_{\rm H^+}&-n_e&0&n_{\rm He^+}&2n_{\rm He^{2+}}\\
0&0&0&1&-1&1&0\\
0&0&0&1&0&-1&1\\
0&0&0&0&n_{\rm He}&n_{\rm He^+}&n_{\rm He^{2+}}
\end{array}
\hspace{-1ex}\right)
\hspace{-1ex}\left(\hspace{-1ex}
\begin{array}{c}
d\ln n_{\rm H_2}\\
d\ln n_{\rm H}\\
d\ln n_{\rm H^+}\\
d\ln n_{\rm He}\\
d\ln n_{\rm He^+}\\
d\ln n_{\rm He^{2+}}\\
d\ln n_e
\end{array}\hspace{-1ex}\right)
\hspace{-1ex}=\hspace{-1ex}\left(\hspace{-1ex}
\begin{array}{c}
-d\ln K_{\rm dis}\\
n_{\rm tot}^{\rm H}d\ln n_{\rm tot}^{\rm H}\\
d\ln K_{\rm ion}\\
0\\
d\ln K_{\rm He,1}\\
d\ln K_{\rm He,2}\\
n_{\rm tot}^{\rm He}d\ln n_{\rm tot}^{\rm He}
\end{array}\hspace{-1ex}\right).
\hspace{1em}
\end{eqnarray}
From this matrix equation we can numerically derive the required derivatives such as $\left(\frac{\partial\ln n_i}{\partial\ln T}\right)_\rho$ and $\left(\frac{\partial\ln n_i}{\partial\ln \rho}\right)_T$. For $\left(\frac{\partial\ln n_i}{\partial\ln T}\right)_\rho$ the RHS vector becomes 
\begin{eqnarray}
\left(
\begin{array}{c}
\frac{d\ln z_{\rm H_2}}{d\ln T}-2\frac{d\ln z_{\rm H}}{d\ln T}\\
0\\
\frac{d\ln z_{\rm H^+}}{d\ln T}+\frac{d\ln z_e}{d\ln T}-\frac{d\ln z_{\rm H}}{d\ln T}\\
0\\
\frac{d\ln z_{\rm He^+}}{d\ln T}+\frac{d\ln z_e}{d\ln T}-\frac{d\ln z_{\rm He}}{d\ln T}\\
\frac{d\ln z_{\rm He^{2+}}}{d\ln T}+\frac{d\ln z_e}{d\ln T}-\frac{d\ln z_{\rm He^+}}{d\ln T}\\
0
\end{array}\right),
\end{eqnarray}
and for $\left(\frac{\partial \ln n_i}{\partial \ln \rho}\right)_T$
\begin{eqnarray}
\left(
\begin{array}{c}
0\\
n_{\rm tot}^{\rm H}\\
0\\
0\\
0\\
0\\
n_{\rm tot}^{\rm He}
\end{array}\right).
\end{eqnarray}
The temperature derivatives of the partition functions can be calculated analytically.

The total number density is 
\begin{eqnarray}
n=\sum_i n_i,
\end{eqnarray}
and its derivatives are
\begin{eqnarray}
\left(\frac{\partial \ln n}{\partial \ln T}\right)_\rho\equiv n_T=\sum_i \frac{n_i}{n}\left(\frac{\partial \ln n_i}{\partial \ln T}\right)_\rho,\\
\left(\frac{\partial \ln n}{\partial \ln \rho}\right)_T\equiv n_\rho=\sum_i \frac{n_i}{n}\left(\frac{\partial \ln n_i}{\partial \ln \rho}\right)_T.
\end{eqnarray}

\subsection*{Thermodynamic Variables}
We use the relation valid in ideal gas:
\begin{eqnarray}
P=nkT\hspace{0.5em}\left(=\frac{\rho}{\mu m_{\rm H}}kT\right),
\end{eqnarray}
where $\mu=\frac{\rho}{nm_{\rm H}}$ is the mean molecular weight. The derivatives of the pressure are:
\begin{eqnarray}
\left(\frac{\partial \ln P}{\partial \ln T}\right)_\rho\equiv P_T&=&1+n_T,\\ 
\left(\frac{\partial \ln P}{\partial \ln \rho}\right)_T\equiv P_\rho&=&n_\rho.
\end{eqnarray}
The specific entropy of each species is derived from the grand potential:
\begin{eqnarray}
S_i=-\frac{1}{\rho V}\left(\frac{\partial \Omega_i}{\partial T}\right)_{V,\mu_i}=\frac{kn_i}{\rho}\left(1+\frac{d\ln z_i}{d \ln T}-\frac{\mu_i}{kT}\right),
\end{eqnarray}
and its derivatives are:
\begin{eqnarray}
\left(\frac{\partial S_i}{\partial T}\right)_\rho&=&\frac{kn_i}{\rho T}\left[\frac{d^2\ln z_i}{d \ln T^2}+\left\{1+\left(\frac{\partial \ln n_i}{\partial \ln T}\right)_\rho\right\}\frac{d\ln z_i}{d \ln T}\right]-\frac{\partial n_i}{\partial T}\frac{\mu_i}{\rho T},\\
\left(\frac{\partial S_i}{\partial \rho}\right)_T&=&\frac{kn_i}{\rho^2}\left[\left\{\left(\frac{\partial \ln n_i}{\partial \ln \rho}\right)_T-1\right\}\frac{d\ln z_i}{d\ln T}-1\right]+\left(\frac{n_i}{\rho}-\frac{\partial n_i}{\partial \rho}\right)\frac{\mu_i}{\rho T}.\hspace{2em}
\end{eqnarray}
Because the last terms in these derivatives are canceled out by taking summation of species when the chemical reactions are in equilibrium, we can omit these terms. Then the total entropy and its derivatives are defined as:
\begin{eqnarray}
S&=&\sum_i S_i,\\
\left(\frac{\partial S}{\partial T}\right)_\rho\equiv S_T &=&\sum_i \left(\frac{\partial S_i}{\partial T}\right)_\rho,\\
\left(\frac{\partial S}{\partial \rho}\right)_T\equiv S_\rho &=&\sum_i \left(\frac{\partial S_i}{\partial \rho}\right)_T.
\end{eqnarray}

Now we can derive thermodynamic properties we require in radiation hydrodynamic simulations. \\The isothermal and adiabatic sound speeds:
\begin{eqnarray}
c_T\equiv\sqrt{\left(\frac{\partial P}{\partial \rho}\right)_T}&=&\sqrt{\frac{P}{\rho}P_\rho},\\
c_S\equiv\sqrt{\left(\frac{\partial P}{\partial \rho}\right)_S}&=&\sqrt{\frac{P}{\rho}P_\rho-\frac{P}{T}\frac{P_T S_\rho}{S_T}}.
\end{eqnarray}
The adiabatic index:
\begin{eqnarray}
\Gamma=\left(\frac{d\ln P}{d\ln \rho}\right)_S=\frac{\rho}{P}c_S^2.
\end{eqnarray}
The internal energy per volume and its derivative:
\begin{eqnarray}
e_g&=&\sum_i n_i kT\frac{d\ln z_i}{d\ln T},\\
\left(\frac{\partial e_g}{\partial T}\right)_\rho&=&C_V=\rho T S_T.
\end{eqnarray}
Note that we do not use the relation $e=C_VT$ which is valid only in the completely ideal cases, as \citet{boley07} suggested (see also \citet{bb75}).

We tabularize these thermodynamic variables as functions of $(\rho, e_g)$ and $(\rho, T)$ with sufficiently high resolution ($\Delta \log \rho = 0.05$, $\Delta \log e_g = 0.025$ and $\Delta \log T = 0.02$) in $\rho=10^{-22}$ -- $10 \, {\rm g \, cm^{-3}}$ and $T=3$ -- $10^6 \, {\rm K}$. We use this EOS table with bi-log-linear interpolation.

\subsection*{Comments on EOS}
Our treatment of the EOS for hydrogen and helium is valid in relatively low-density regions ranging from interstellar gas to the second collapse phase. However, in very dense region like the deep interior of the protostellar core, non-ideal effects are not negligible: interactions between particles, weak quantum effects in low-temperature but high-density region, pressure ionization of hydrogen, and contributions from other chemical species. Such non-ideal effects will affect the thermodynamics and the dynamics (e.g., the quasi-equilibrium state of the second core may vary). Actually, our EOS results in a serious unphysical behavior in the very high density region ($\rho > 0.1 \,{\rm g\, cm^{-3}}$) that almost all the hydrogen particles are turned into the molecular form even when the gas temperature is high enough to destruct the hydrogen molecule. This is because of the assumption of the ideal chemical equilibrium, but in reality this assumption broke down there due to the neglected interactions between particles \citep{scv95}. Then our EOS gives the considerably soft adiabatic index $\Gamma$ (Figure~\ref{s_ee}) because of the contribution from the vibration transitions of molecular hydrogen. Since this behavior is completely unphysical, our EOS is invalid in such a high density region. More realistic EOS such as \citet{scv95} is required to calculate this region properly.

\section*{Appendix 2. Opacities}
For the gray radiation transfer, we use the compiled tables of the Rosseland and Planck mean opacities of \citet{semenov}\footnote{http://www.mpia.de/homes/henning/Dust\_opacities/Opacities/opacities.html}, \citet{fer05}\footnote{http://webs.wichita.edu/physics/opacity/} and the Opacity Project (OP) \citep{op94}\footnote{http://cdsweb.u-strasbg.fr/topbase/TheOP.html}. For dust opacities, we adopt the composite aggregate dust model of the normal abundance from Semenov's mean opacity tables. Though Semenov's tables also contain the gas opacities, we adopt the gas opacities of \citet{fer05} and \citet{op94} because Semenov's Planck mean opacity seems to be significantly lower than other opacity tables \citep{fer05,op94} where molecular and atomic lines dominate the opacity sources. Therefore we connect the tables of Semenov et al. (2003) and Ferguson et al. (2005) at the temperature where all the dust components evaporate. The dust evaporation temperature (it weakly depends on the gas density, but in the typical density region, $T\sim 1400$ -- $1500\, {\rm K}$) is given in Semenov's opacity calculation code. We use \citet{fer05} in the low temperature region ($\log T<4.5$) and OP in the high temperature region ($4.5 < \log T < 6$). We tabulate these tables as functions of ($\rho,T$) and use them with bi-log-linear interpolation.

\begin{figure}[tb]
\begin{center}
\scalebox{1.2}{\includegraphics{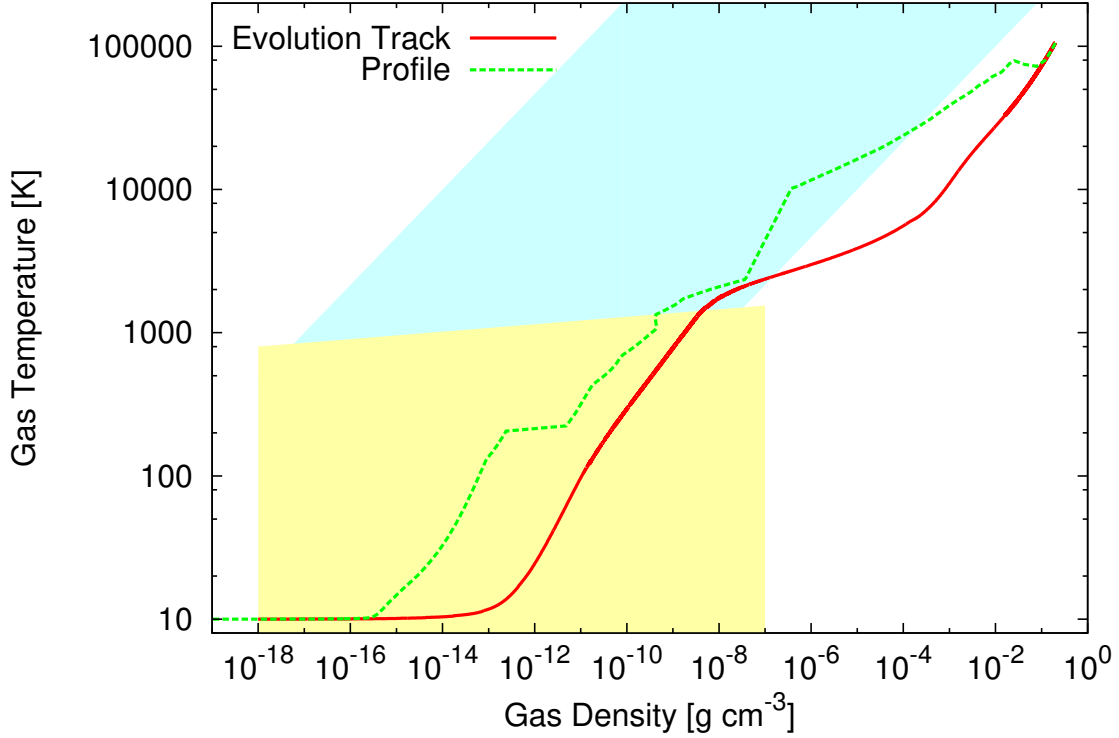}}
\caption{The blue shaded region is covered by gas opacity tables \citep{op94,fer05} and the yellow region is covered by dust opacity tables \citep{semenov}. The border line between blue and yellow corresponds to the dust evaporation temperature. The red line represents the typical evolution track of the central gas element in the spherical protostellar collapse and the green line does the profile at the end of the simulation.}
\label{opcov}
\end{center}
\end{figure}

\begin{figure}[htb]
\begin{center}
\scalebox{0.4}{\includegraphics{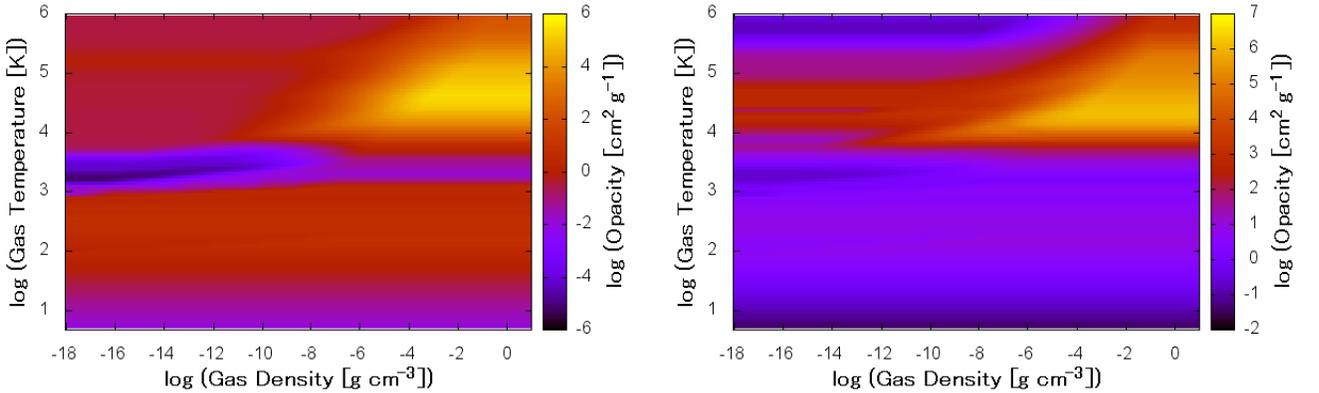}}
\caption{The Rosseland (left) and Planck (right) mean opacities.}
\label{opacity}
\end{center}
\end{figure}

Unfortunately, the opacity tables do not cover the whole required region. Figure~\ref{opcov} shows the coverage of the opacity tables in the $\rho-T$ plane. The typical evolution track of the central gas element and the profile in the spherically symmetric collapse are also plotted. The dust opacities of \citet{semenov} cover $10^{-18}\,< \rho \,({\rm g\, cm^{-3}}) <10^{-7}$. OP and \citet{fer05} cover $-8<\log R<1$ where $R=\rho/T_6^3$ and $T_6=T \,{\rm (K)}/10^6$. It is not serious that the very low density region is not covered because that region is extremely optically thin and the details of dust opacities do not matter there. We simply extrapolate the opacities by taking the nearest value at the given temperature. The high density region is far more problematic; we do not have proper opacities for the protostellar core. But actually the thermal evolution in this region is dominated by chemical reactions (dissociation and ionization) and radiation transfer is of less importance there because the gas is extremely optically thick. Therefore we dare to extrapolate the tables in the same manner\footnote{We must be careful, however, that this problem becomes more serious when we calculate the evolution of the protostar longer than the Kelvin-Helmholtz timescale before the onset of convection (or deuterium burning). As we can see from Figure~\ref{opcov}, the earliest phase of the protostar is convectively stable (see also \citet{sst1,sst2}) and therefore radiation plays a critical role in heat transfer. More elaborate opacity tables covering wider region are highly demanded.}.

We show the combined opacity tables in Figure~\ref{opacity} as functions of $\rho$ and $T$. Note that there is quite large uncertainty in opacities, especially due to the dust models such as the structure, composition, size distribution and so on. The thermal evolution and dynamics in star formation remain qualitatively similar even when we change the dust parameters, but the observational properties such as Spectral Energy Distributions (SED) are directly affected by the differences between the (monochromatic) opacities.

\section*{Appendix 3. Resistivity}
We calculate Ohmic resistivity considering both thermal and non-thermal particles. For non-thermal processes, we adopt the table of resistivity based on the reaction network model between dust grains and chemical species constructed by \citet{un09} (see also \citet{okz09}). The table of $\eta_{NT}$ is given as a function of the gas density $\rho$, the temperature $T$ and the ionization rate $\xi$. Here we assume the typical interstellar ionization rate due to the cosmic rays, $\xi_{CR}\sim 10^{-17} \, {\rm s^{-1}}$, neglecting shielding by the gas. Since the attenuation depth of the cosmic rays is about $100\, {\rm g\, cm^{-2}}$, the ionization rate will be lower in the deep interior of the first core. In this sense, and also because we neglect ambipolar diffusion, our models corresponds to the lower limit of (but still highly efficient) magnetic flux loss. Because decay of radionuclides (${\rm ^{26}Al}$ is the dominant source) considerably contributes to the ionization rate, $\xi_{RA}\sim 7.6 \times 10^{-19} \, {\rm s^{-1}}$ \citep{un09}, the effect from neglecting the shielding is as large as about an order of magnitude at most. We should mention that there is still large uncertainty in the resistivity from the grain properties such as the structure, composition, size distribution and so on.

In order to calculate the resistivity for our simulations, it is sufficient to estimate the contribution to the thermal ionization from the species which has low ionization energy. Potassium (K) has very low ionization energy ($kT_{ion}\sim 4.33\,{\rm eV}$) and sufficiently abundant to recover good coupling between gas and magnetic fields, therefore it is the most important electron-supplying species in star formation processes. Here we calculate the resistivity due to the thermal ionization of potassium on the assumption of thermal equilibrium using the following equation:
\begin{eqnarray}
\eta_T =7.5\times 10^9 \exp\left(\frac{25188\,{\rm K}}{T}\right)T^{-1/4} \rho^{1/2} \hspace{1em} \, \rm{cm^2 \, s^{-1}}.
\end{eqnarray}
We calculate the total resistivity as follows:
\begin{eqnarray}
\eta^{-1}=\eta_T^{-1}+\eta_{NT}^{-1},
\end{eqnarray}
because the resistivity is inversely proportional to the ionization degree.

\begin{figure}[tb]
\begin{center}
\scalebox{1}{\includegraphics{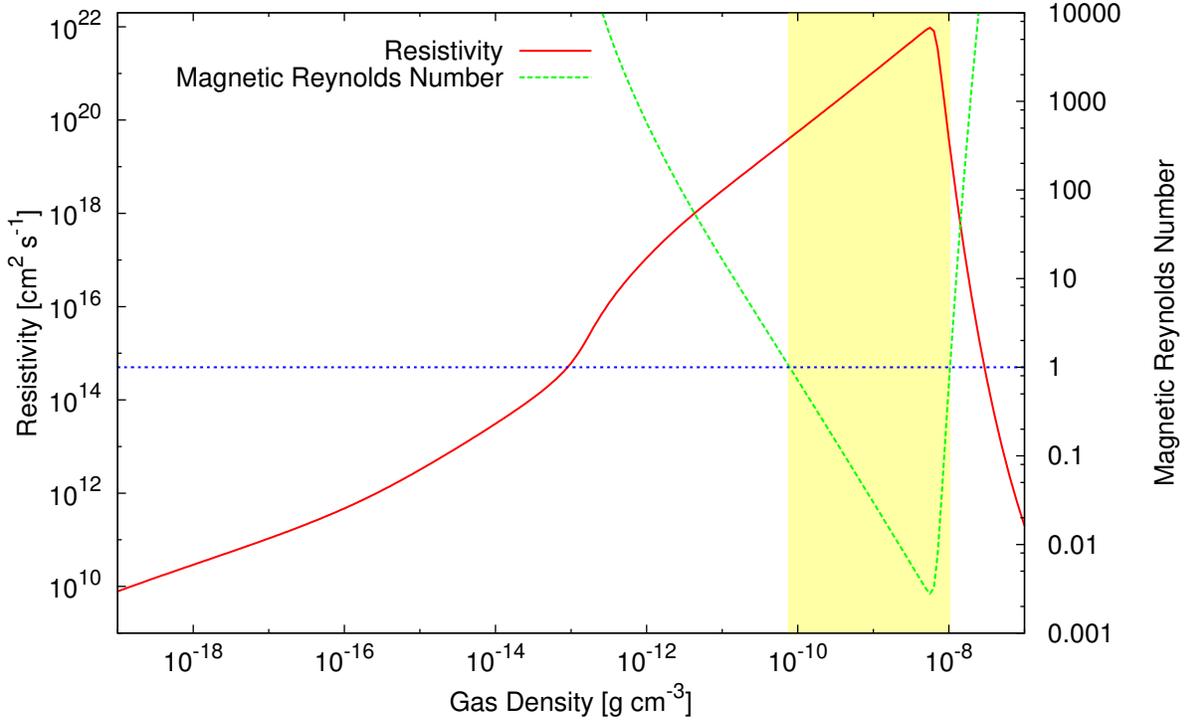}}
\caption{The resistivity $\eta$ and the magnetic Reynolds number $R_m$ are plotted as functions of the gas density. Magnetic fields are decoupled from fluid where $R_m < 1$ (yellow shaded region).}
\label{resist}
\end{center}
\end{figure}

We show the resistivity and magnetic Reynolds number $R_m\equiv v_{\rm ff}\lambda_{\rm J}/\eta$ as a function of the gas density in Figure~\ref{resist}. Here we consider the Jeans length $\lambda_{\rm J}\equiv 2\pi c_s \sqrt{\frac{3\pi}{32G\rho}}$ and the free-fall velocity $v_{\rm ff}\equiv \sqrt{\frac{GM_{\rm J}}{\lambda_{\rm J}}}$ (where $M_{\rm J}=\frac{4\pi}{3}\lambda_{\rm J}^3\rho$) as typical length and velocity scales to estimate the magnetic Reynolds number. To draw this plot, we adopt the barotropic approximation as a typical thermal evolution to calculate the gas temperature:
\begin{eqnarray}
T=\max\left[10, 10\times\left(\frac{\rho}{\rho_{crit}}\right)^{\Gamma-1}\right] \, {\rm K},
\end{eqnarray}
where $\rho_{crit}=2\times 10^{-13} \, {\rm g\,  cm^{-3}}$ is the critical density and $\Gamma=7/5$ is the adiabatic index. The resistivity steeply decreases in $\rho \gtsim 10^{-8} \, {\rm g\, cm^{-3}}$ because of the thermal ionization of potassium. The magnetic fields are decoupled from fluid where the magnetic Reynolds number is less than unity. Our resistivity is significantly lower than the resistivity of \citet{kunz09} (see also \citet{kunz10,dapp12}). This difference mainly comes from the ionization rates; they assume that decay of ${\rm ^{40}K}$ ($\xi_{\rm K}\sim 2.43 \times 10^{-23} \, {\rm s^{-1}}$) is the dominant ionization source in the dense region where cosmic rays cannot penetrate, and neglect the contribution from short-lived species like ${\rm ^{26}Al}$. Our resistivity is quite similar to that used in \citet{mim06}.

\end{document}